\documentclass[twocolumn,final,10pt]{asme2ej}
\pdfoutput=1
\usepackage[mathscr]{euscript}
\usepackage{algorithm,algorithmic}
\usepackage{amsmath,amssymb,amsfonts} 
\usepackage{bm}
\usepackage{graphicx}
\graphicspath{ {images/} }
\usepackage{enumitem}

\setlist[itemize,1]{label={\fontfamily{cmr}\fontencoding{T1}\selectfont\textbullet}}

\DeclareFixedFont{\ttb}{T1}{txtt}{bx}{n}{12} 
\DeclareFixedFont{\ttm}{T1}{txtt}{m}{n}{12}  

\renewcommand{\vec}[1]{\mathbf{{#1}}}
\newcommand{\Rq}{R_\text{qoi}}

\newcommand{\Nq}{{N_\text{quad}}}
\newcommand{\Nkl}{{N_{kl}}}
\newcommand{\Nkla}{{N_{kl}^a}}
\newcommand{\Nklp}{{N_{kl}^p}}

\newcommand{\var}{\mathrm{Var}}

\title{Model input and output dimension reduction using Karhunen Lo\`{e}ve 
expansions with application to biotransport}
\author{Alen Alexanderian
        \affiliation{
	Department of Mathematics\\
	North Carolina State University\\
	Raleigh, NC 27695\\
        Email: alexanderian@ncsu.edu}}
\author{William Reese
    \affiliation{
	Department of Mathematics\\
	North Carolina State University\\
	Raleigh, NC 27695}}	
\author{Ralph C.~Smith
    \affiliation{
	Department of Mathematics\\
	North Carolina State University\\
	Raleigh, NC 27695}}
\author{Meilin Yu\thanks{Address all correspondence to this author.}
    \affiliation{Department of Mechanical Engineering\\
    	University of Maryland, Baltimore County\\
    	Baltimore, MD 21250\\
        Email: mlyu@umbc.edu}}

\begin{document}
\setlength{\abovedisplayskip}{4pt}
\setlength{\belowdisplayskip}{4pt}

\maketitle    

\begin{abstract}
\textit{
We consider biotransport in tumors with uncertain heterogeneous material
properties.  Specifically, we focus on the elliptic partial differential
equation (PDE) modeling the pressure field inside the tumor.  
The permeability field is modeled as a log-Gaussian random field with a
prespecified covariance function. 
%
%
We numerically explore dimension reduction of the input parameter and model
output. 
Truncated Karhunen--Lo\`{e}ve (KL) expansions are used to decompose
the log-permeability field, as well as the resulting random pressure field.
We find that although very high-dimensional representations are needed to
accurately represent the permeability field, especially in presence of small
correlation lengths, the pressure field is not very sensitive to
high-order KL terms of the input parameter.  
Moreover, we find that the pressure field itself can be represented accurately 
using a KL expansion with a small number of terms. 
These observations are used to guide a reduced-order modeling approach 
to accelerate computational studies of biotransport in tumors.
}
\end{abstract}



\section{INTRODUCTION}

We focus on modeling biotransport processes in tumors with uncertain
heterogeneous material properties. An improved understanding of these  processes
can provide vital insight for agent delivery in cancer
treatment~\cite{maher:08,Deb:09CPD}.  Biotransport processes in tumors can be
modeled as flows in heterogeneous porous media. Equations governing
biotransport consist of an elliptic PDE describing the pressure distribution
and a hyperbolic PDE that describes agent (e.g., drug) delivery in porous
media~\cite{Swartz:07ARBE}.  The uncertain tumor material properties can be
modeled as random fields, which are then incorporated as coefficient functions
in the governing PDEs.

%
%

In the present work, the uncertain permeability field is modeled as a
log-Gaussian random field. Our aim is to efficiently simulate the uncertain
pressure field.  
We use KL expansions~\cite{Loeve77,Ghanem} to represent the random
log-permeability field.
The use of KL expansions for representing random field
parameters in mathematical models has been a common modeling approach in the
uncertainty quantification community~\cite{Ghanem98,LeMaitreReaganNajmEtAl02,XiuKarniadakis03,
LeMaitreKnio04,BabuvskaNobileTempone07,
Doostan07,SaadGhanem09,MatthiesKeese05,Graham15KuoKuoNicholsEtAl15,Elman17}.  
The typical approach is to use a truncated KL expansion with enough term to ensure
the average variance of the parameter field is sufficiently captured. 
That is, the truncation of the KL expansion is performed a priori and without
taking the response of the governing equations to the random field coefficients
in mind. 
We take a  
\emph{goal-oriented} point of view: instead of relying on a
truncated KL expansion of the log-permeability field that is computed
independently of the governing PDE, we seek to retain only the KL terms that
the PDE solution operator is sensitive to.  This goal-oriented strategy can
lead to significant input parameter dimension reduction, especially for input
fields with small correlation lengths.  The PDE solution---the pressure
field---itself can also be represented via a truncated KL expansion.  We
observe that a low-rank representation of the pressure field is often afforded
by a truncated KL expansion with a small number of terms. The latter is a
consequence of the (often) rapid decay of the eignvalues of the model output
covariance operator.  Our approach guides an input and output dimension
reduction strategy: a low-rank representation of the pressure field can be
computed in a low-dimensional parameter space. We mention that a preliminary
version of this work was presented in the conference
paper~\cite{AlexanderianReeseSmithEtAl18}.

This article is structured as follows.
In Section~\ref{sec:background}, we recall the requisite background
material on random fields and their KL expansion. In  that section, 
we also outline a computational strategy for computing KL expansions
for random fields with or without a prespefied covariance function.
In Section~\ref{sec:1D}, we use a model elliptic PDE in one space dimension 
to illustrate the components of the proposed approach. Then, in 
Section~\ref{sec:biotransport}, we focus on a biotransport application
problem. We present numerical results illustrating the merits of the proposed
strategy. Concluding remarks are provided in Section~\ref{sec:conc}.

\section{BACKGROUND ON RANDOM FIELDS}\label{sec:background}
Let $(\Omega, \mathcal{F}, P)$ be a probability space, where 
$\Omega$ is a sample space, $\mathcal{F}$ is an appropriate $\sigma$-algebra, 
and $P$ is a probability measure.
Let $X \subset \mathbb{R}^d$, with $d =1, 2$, or $3$, be a
compact set. 
Let $Z : X \times \Omega \rightarrow
\mathbb{R}$ be a stochastic process~\cite{Williams}.  
From a modeling standpoint, $Z(\vec{x}, \omega)$ can be used to
represent uncertain parameters fields in mathematical models. 

A stochastic process is called \emph{centered} 
if $E[Z(\vec{x}, \cdot)] = 0$ for all $\vec{x} \in X$, 
where $E[Z(\vec{x}, \cdot)] = \int_\Omega Z(\vec{x}, \omega) \, P(d\omega)$.
A process $Z$ is called 
\emph{mean square continuous} if 
\[
\lim_{\vec{h} \to \vec{0}} E[\left(Z(\vec{x}+\vec{h}, \cdot) - Z(\vec{x}, \cdot)
\right)^{2}] = 0, \quad \text{for all } \vec{x} \in X.
\]

The covariance function $c : X \times X \rightarrow \mathbb{R}$ and
the corresponding correlation function of a stochastic
process $Z$ are, respectively, given by  
\begin{equation*}
\begin{aligned}
c(\vec{x},\vec{y}) &= E[Z(\vec{x},\cdot) Z(\vec{y}, \cdot)] 
   - E[Z(\vec{x}, \cdot)]E[Z(\vec{y}, \cdot)],
\\ 
   \rho(\vec{x}, \vec{y}) &= \frac{c(\vec{x}, \vec{y})}
      {\sqrt{c(\vec{x},\vec{x})} \sqrt{c(\vec{y},\vec{y})}}. 
\end{aligned}
\end{equation*}
We also recall the definition of the \emph{covariance operator} of 
a stochastic process $Z(\vec{x}, \omega)$, which is given by 
\begin{equation}~\label{cov_operator}
   [Cu](\vec{x}) = \int_{X} c(\vec{x},\vec{y}) u(\vec{y}) d\vec{y}, 
   \quad u \in L^2(X).
\end{equation}

\textbf{Karhunen--Lo\`{e}ve expansion.} 
Let $Z: X \times \Omega \rightarrow \mathbb{R}$ be a centered mean-square continuous 
stochastic process, and let $\{ e_i \}_{i=1}^\infty$ be the orthonormal 
basis of eigenvectors of its covariance operator with
corresponding
(non-negative) eigenvalues $\{\lambda_i\}_{i=1}^\infty$:
\begin{equation}\label{equ:eigenvalue}
    \int_X c(\cdot, \vec{y}) e_i(\vec{y}) \, d\vec{y} = \lambda_i e_i(\cdot), \quad i =1, 2, \ldots.
\end{equation}
The process $Z(\vec{x}, \omega)$ can be represented via its
KL expansion~\cite{Loeve77,Ghanem,knio,Smith13}: 
	\begin{equation}\label{equ:KLE}
	Z(\vec{x}, \omega) = \sum_{i = 1}^{\infty} \sqrt{\lambda_i} \xi_{i}(\omega) e_{i}(\vec{x}), 
	\end{equation}
where $\xi_{i}$ are centered mutually uncorrelated random variables with unit
variance and are defined by
\begin{equation*}
\xi_{i}(\omega) = \frac{1}{\sqrt{\lambda_{i}}} \int_{X} Z(\vec{x},\omega) e_{i}(\vec{x}) d\vec{x}.
\end{equation*}
The convergence of the series~\eqref{equ:KLE} is uniform in $X$, and is mean
square in $\Omega$~\cite{Loeve77}.  
Moreover, if $Z(\vec{x},\omega)$ is a Gaussian process, convergence of the series
~\eqref{equ:KLE} is almost sure for each $\vec{x} \in X$; 
see~\cite[p.~485]{Loeve77} for further details.

\textbf{Numerical computation of KL expansion.} 
To compute the KL expansion of a stochastic process the eigenvalue
problem~\eqref{equ:eigenvalue} must be solved first.  In the present work, we
follow Nystr\"{o}m's approach~\cite{Kress03}, which involves discretizing the
generalized eigenvalue problem using quadrature.  We describe the steps for
computing KL expansions below.  Further details on numerical methods for
computing KL expansions can be found in~\cite{BetzPapaioannouStraub14}.  

When modeling random field
coefficients in models, one often has access to a prespecified covariance
function. On the other hand, when computing KL expansion of a random field
output of a mathematical model we only have access to realizations
of the model output.
Let $U$ denote the random field output of model governed by PDEs. 
In practice, often the model uncertainties are parameterized using 
a random vector $\bm{\xi}$, in which case the random field output 
$U = U(\vec{x}, \bm{\xi})$ can be
computed for specific realizations of $\bm\xi$.  
To compute the truncated KL expansion,
\begin{equation}\label{equ:KLE_U}
\begin{aligned}
U(\vec{x}, \bm{\xi}) &\approx 
\bar{U}(\vec{x}) + \sum_{i=1}^\Nkl \sqrt{\lambda_i} u_i(\bm{\xi}) e_i(\vec{x}), 
\\
\bar{U}(\vec{x}) &= E[U(\vec{x}, \cdot)],
\\
&u_i(\bm{\xi}) = \frac{1}{\sqrt{\lambda_i}} 
\int_{X} (U(\vec{x},\bm{\xi}) - \bar{U}(\vec{x}))e_{i}(\vec{x}) d\vec{x},
\end{aligned}
\end{equation}
the covariance function of $U$ needs to be approximated
via sampling, resulting in an approximate covariance operator 
$C$ for the process.  
Then, the generalized eigenvalue problem will be solved using this
approximate covariance operator to find (approximations to) $\lambda_i$ and
$e_i$, $i = 1, \ldots, \Nkl$.  In practice, the dominant KL terms can be
captured reliably, with a modest sample size, as discussed in our numerical
results. 
We summarize the steps required for computing truncated KL expansion of 
the random process $U(\vec{x}, \bm\xi)$ in
Algorithm~\ref{alg:KLE}.
In what follows, we refer to the coefficients $u_i$ 
in~\eqref{equ:KLE_U} as the \emph{KL modes}. 

\begin{algorithm*}[h]
\renewcommand{\algorithmicrequire}{\textbf{Input:}}
\renewcommand{\algorithmicensure}{\textbf{Output:}}
\caption{Computing KL expansion of a random process $U(\vec{x}, \bm{\xi})$ 
using Nystr\"{o}m's 
approach.}
\label{alg:KLE}
\begin{algorithmic}[1]
\REQUIRE 
(i) A quadrature formula on $X$ with nodes and weights $\{ \vec{x}_m, w_m \}_{m = 1}^\Nq$;
(ii) function evaluations $\{U(\vec{x}_m, \bm\xi^k)\}$, $m \in \{1, \ldots, \Nq\}$, 
     $k \in \{1, \ldots, N\}$; (iii) trunction level $\Nkl$. 

\ENSURE Eigenpairs of the discretized covariace operator, 
$\{(\lambda_i, \vec{e}_i)\}_{i=1}^\Nkl$,
and KL modes $\{u_i\}_{i=1}^\Nkl$. 
\STATE Compute the mean 
\[
\bar{U}_m = \frac1N\sum_{j=1}^N U(\vec{x}_m, \bm\xi^j),
\quad m \in \{1, \ldots, \Nq\}.
\]
\STATE Center the process
\vspace{-2mm}
\[
    u_c(\vec{x}_m, \bm\xi^k) = U(\vec{x}_m, \bm\xi^k) - \bar{U}_m, 
    \quad k \in \{1, \ldots, N\}, m \in \{1, \ldots, \Nq\}.
\]
\vspace{-2mm}
\STATE Form the covariance matrix 
\vspace{-2mm}
\[
 K_{lm} =  \frac{1}{N-1} \sum_{k = 1}^{N} u_c(\vec{x}_l, \bm\xi^k) u_c(\vec{x}_m, \bm\xi^k), \quad l,m \in \{1, \ldots, \Nq\}.
\]
\vspace{-2mm}
\STATE Let $\mathbf{W} = \mathrm{diag}(w_1, w_2, \ldots, w_\Nq)$ and solve the eigenvalue problem
\vspace{-2mm}
\[
 \mathbf{W}^{1/2} \mathbf{K} \mathbf{W}^{1/2} \vec{v}_i = \lambda_i \vec{v}_i,
\quad 
 i \in \{ 1, \ldots, \Nq\}. 
\]
\vspace{-4mm}
\STATE Compute $\vec{e}_i = \mathbf{W}^{-1/2} \vec{v}_i$, $i \in \{ 1, \ldots, \Nq\}$.
\STATE Compute the discretized KL modes, 
\vspace{-2mm}
\[
{u}_i(\bm\xi^k) = \frac{1}{\sqrt{\lambda_i}}
\sum_{m = 1}^\Nq w_m u_c(\vec{x}_m, \bm\xi^k) e_i^m, \quad i \in \{1, \ldots, \Nkl\},
\, k \in \{1, \ldots, N\}.
\]
\end{algorithmic}
\end{algorithm*}

\section{MODEL 1D ELLIPTIC EQUATION WITH RANDOM COEFFICIENT FUNCTION}
\label{sec:1D}

We let $(\Omega, \mathcal{F}, P)$ be a
probability space, and for $\omega \in \Omega$, consider  
the following model elliptic boundary value problem: 
\begin{equation}\label{equ:random_model_prob}
\begin{aligned}
-\frac{d}{dx}\left(\kappa(x,\omega)\frac{dp(x,\omega)}{dx}\right) &= f(x), \hspace{.2in} x\in D = (-1, 1),\\
p(-1,\omega) &= 1, \\
p(1,\omega)  &= 0.
\end{aligned}
\end{equation}
In the following numerical experiments, 
the right hand side function is given by $f(x) = \cos(\pi x) + \sin(2\pi x)$.
We model the 
coefficient function $\kappa(x, \omega)$ as a 
log-Gaussian random field as follows.
Let $Z(x, \omega)$ be a centered Gaussian process with covariance function,
\begin{equation}\label{eq:Ornstein-Uhlenbeck_specific}
 c_Z(x,y) = \exp\Big\{\!\!-\!\frac{|x-y|}{\ell}\Big\}.
\end{equation}  
We set the correlation length $\ell$ of the process to 
$\ell = 1/4$.
Then, we set $\kappa(x, \omega) = \exp(a(x, \omega))$ with 
\begin{equation}\label{equ:KL_trunc}
    a(x, \omega) = a_0(x) + \sigma Z(x, \omega),
\end{equation} 
where $a_0$ and $\sigma^2$ are the pointwise mean and variance of $a(x,
\omega)$, respectively;  we choose these parameters such that the pointwise
mean and standard deviation of $\kappa(x, \omega)$ are $m=0.1$ and $s = .07$,
respectively. Accordingly, we 
let
$\sigma^2 = \log(1 + s^2/m^2)$ 
and $a_0 \equiv \log\big(m/\sqrt{1+s^2/m^2}\big)$.~\footnote{
We have used 
the well-known formulas relating the mean and variance of a log-normal random 
variable $Y = \exp(a_0 + \sigma X)$, where $X$ is standard normal, 
to $a_0$ and $\sigma^2$.} 
We consider the weak formulation of the problem~\eqref{equ:random_model_prob}, 
and use the continuous Galerkin finite element method, with linear basis
functions, to solve the problem numerically.

We use a truncated KL expansion for $a(x, \omega)$, 
\begin{equation} \label{equ:aKL}
a_\Nkla(x,\omega) = a_0 + \sigma \sum_{i = 1}^{\Nkla} \sqrt{\lambda_i} \xi_{i}(\omega) e_{i} (x), 
\end{equation}
where $(\lambda_i, e_i)$ are eigenpairs of the covariance operator of $Z(x,
\omega)$. Due to the Gaussianity of the process, $\xi_i$ are independent
standard normal random variables. Note that 
the random vector 
\begin{equation}\label{equ:xi}
\bm{\xi} = \begin{bmatrix}
\xi_1 & \xi_2 & \cdots & \xi_\Nkla
\end{bmatrix}^T
\end{equation}
completely parameterizes the uncertainty in the 
problem~\eqref{equ:random_model_prob}, and
its solution $p({x}, \omega) = p(x, \bm{\xi}(\omega))$.

\begin{figure*}[h]\centering
\includegraphics[width=.35\textwidth]{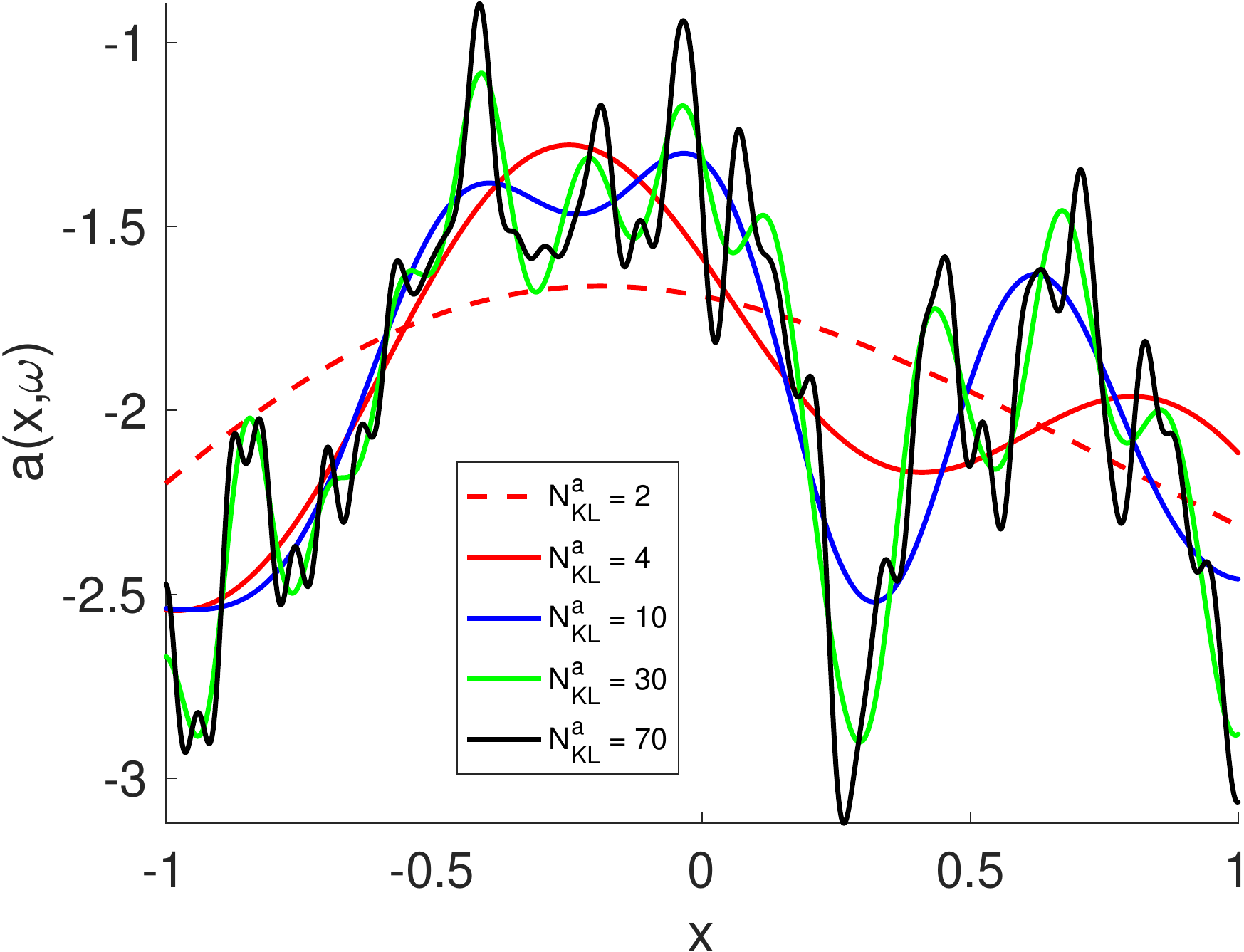}
\includegraphics[width=.35\textwidth]{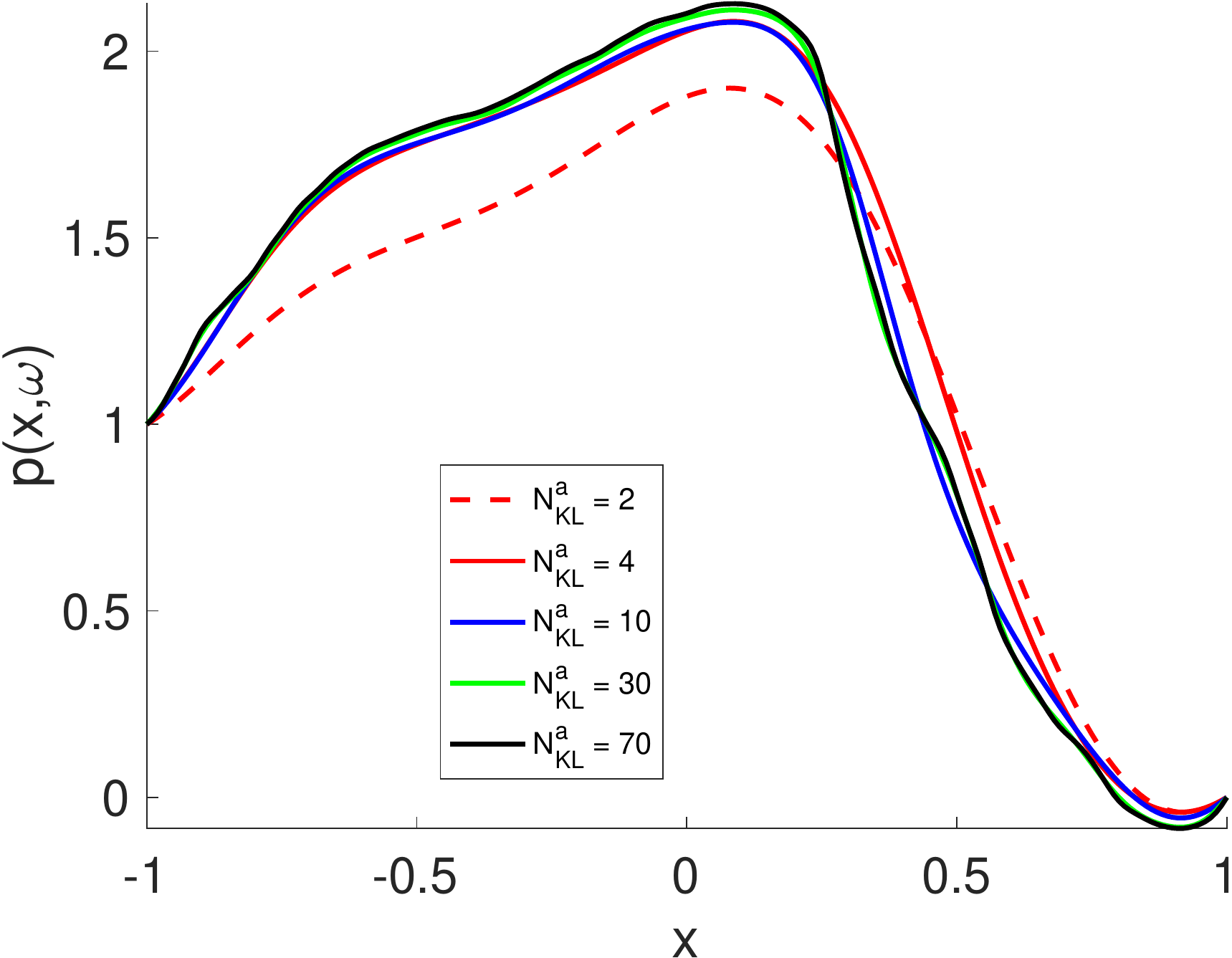}
\caption{
Left: a fixed realization of the log-coefficient field as we increase 
$N_\text{KL}^a$; right: 
Solutions of the problem~\eqref{equ:random_model_prob} corresponding 
a fixed realization of $p(x, \bm\xi)$ as we increase $N_\text{KL}^a$ (right).}
\label{fig:realizations}
\end{figure*}

%
As a first illustration, we consider a fixed realization of $a(x, \omega)$ as
$\Nkla$ increases in Figure~\ref{fig:realizations}~(left). Note that
sufficiently large $\Nkla$ is needed to capture the fluctuations of the random
field.  On the other hand, the corresponding PDE solution is less sensitive
to the higher-order KL terms of the parameter, as seen
in Figure~\ref{fig:realizations}~(right).  This behavior is consistent
with the analysis in~\cite{Cleaves19}, where a global sensitivity analysis
formalism is used to quantify the impact of the KL terms of the
log-coefficient, in an elliptic PDE, on variability in solution of the PDE.


\begin{figure*}[ht]\centering
\begin{tabular}{cc}
\includegraphics[width = .35\textwidth]{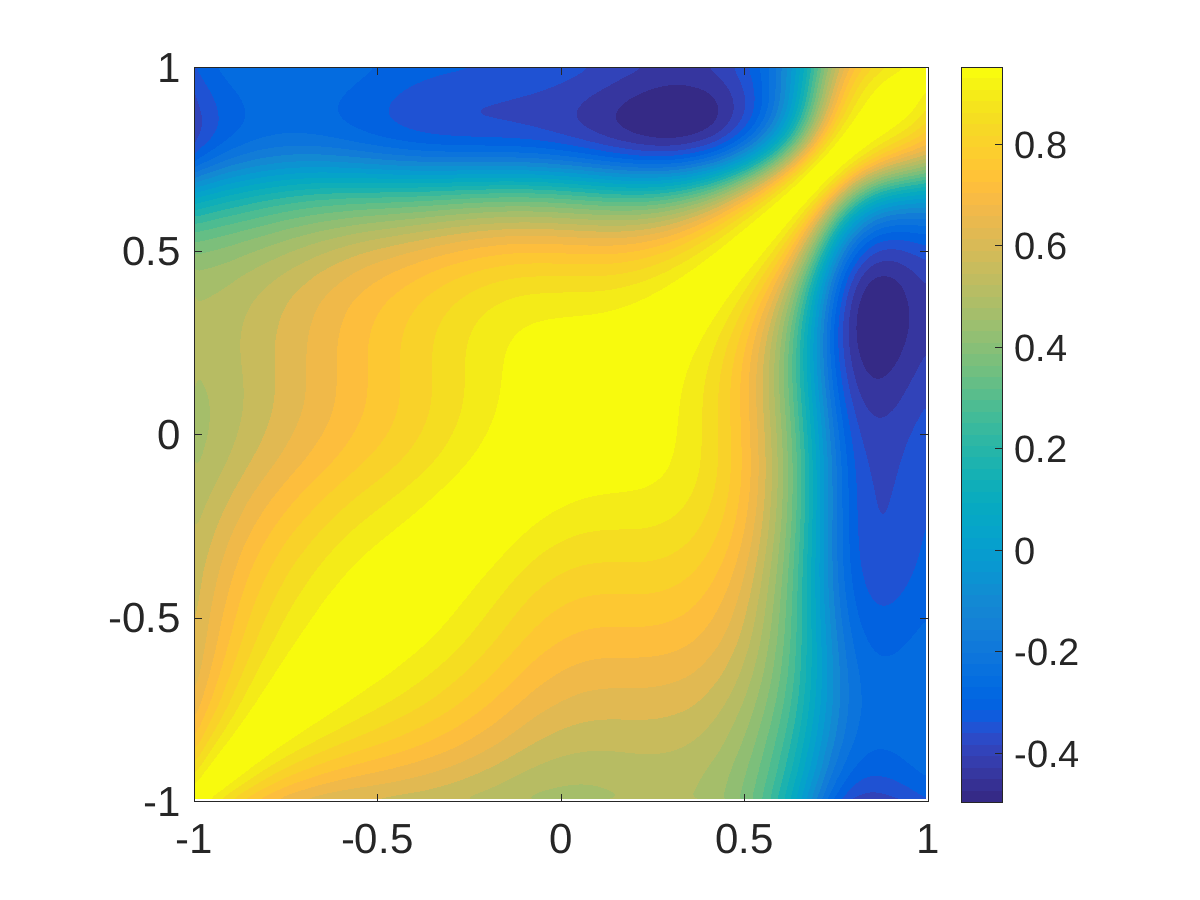}&
\includegraphics[width = .3\textwidth]{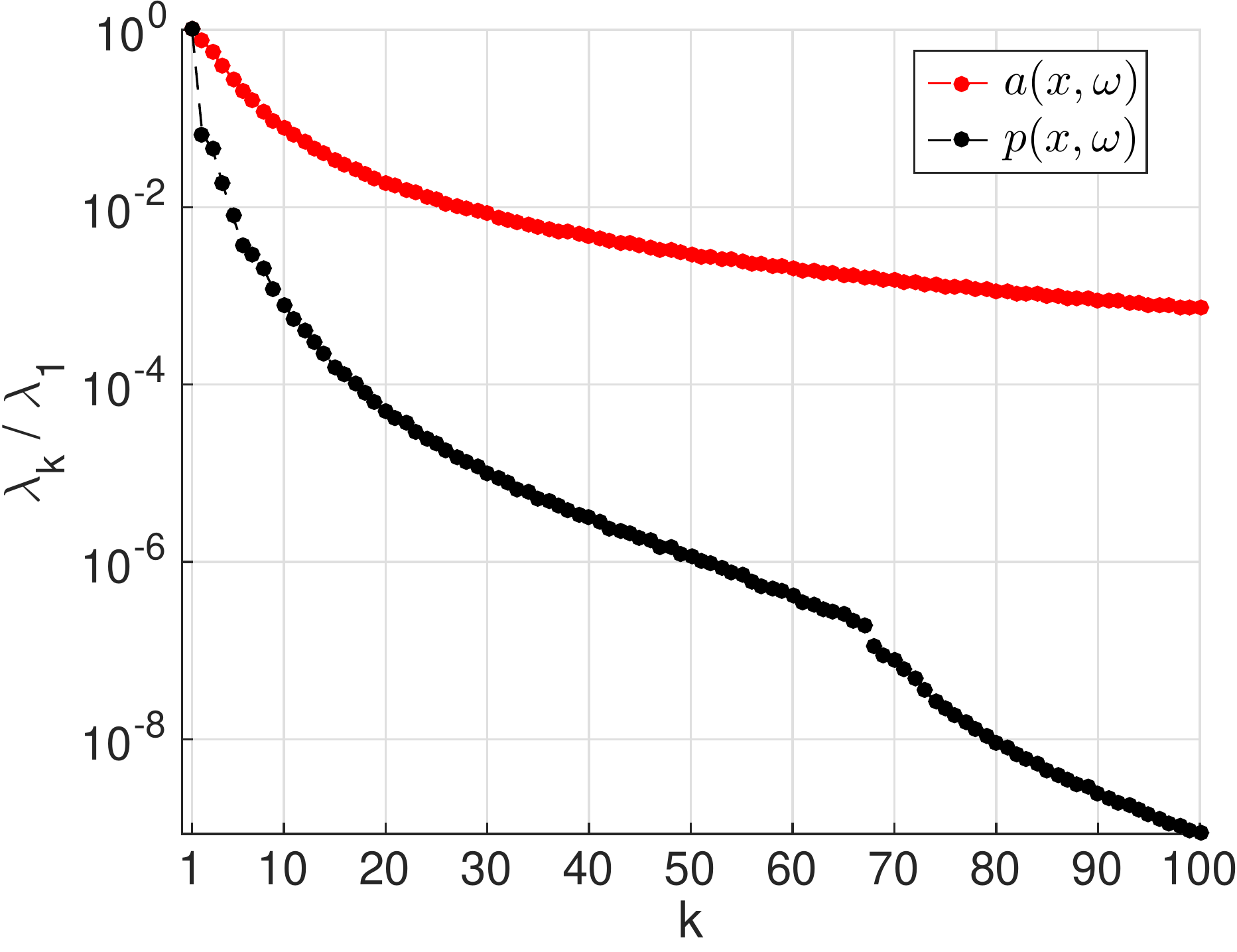}\\
\includegraphics[width = .3\textwidth]{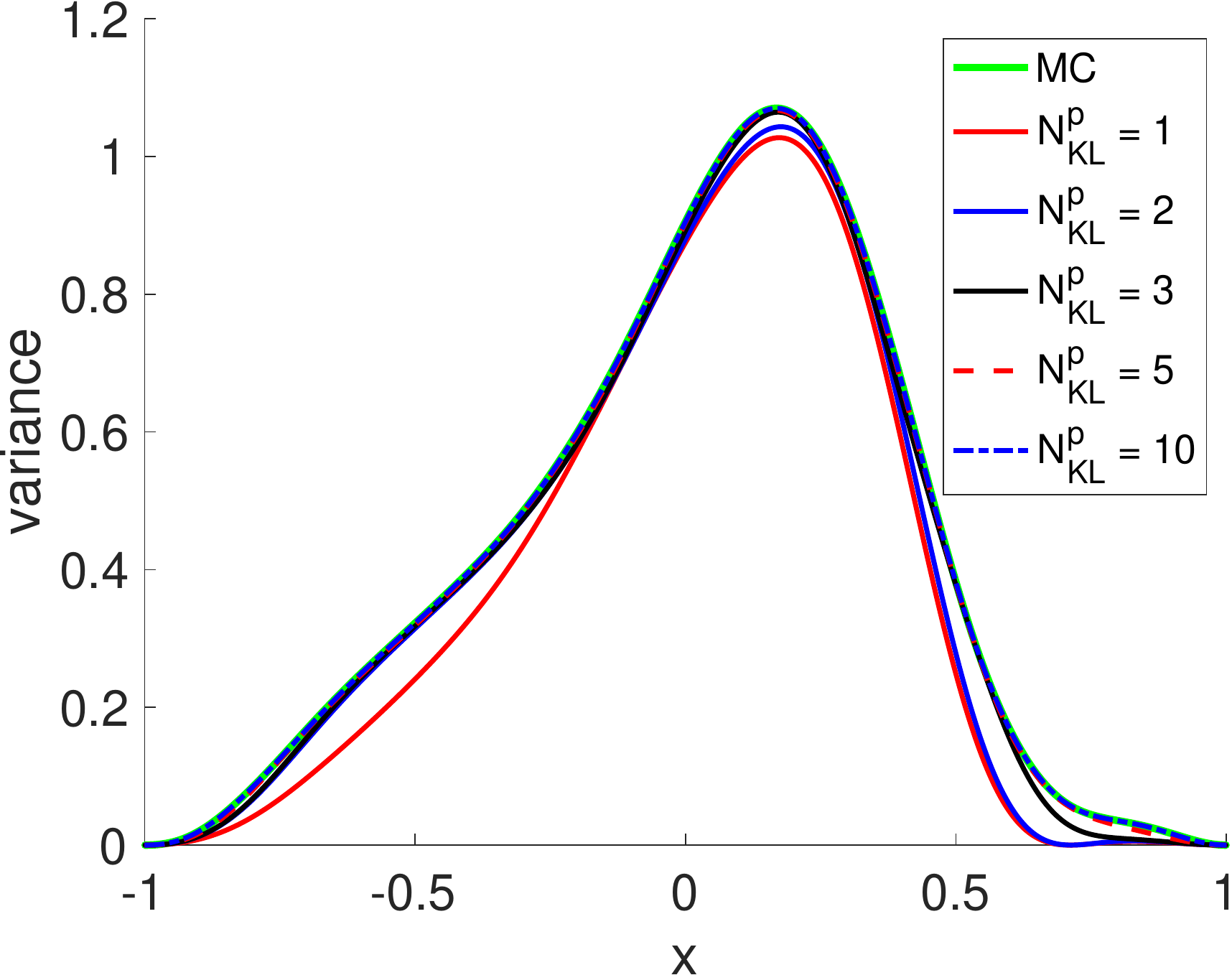}&
\includegraphics[width = .3\textwidth]{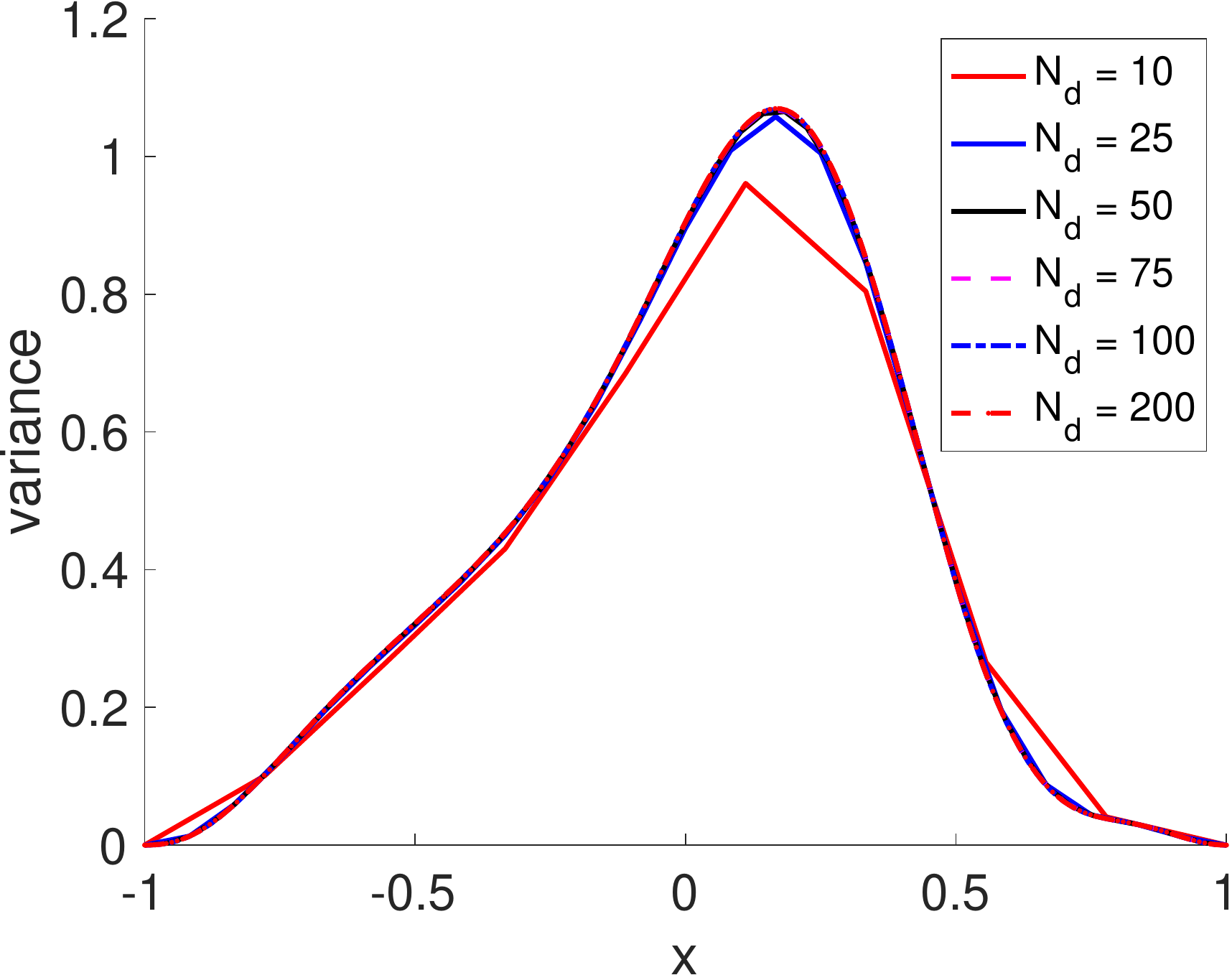}	
\end{tabular}
\caption{Top row: the left image shows 
the correlation function of the output and
the right image depicts the
decay of the spectrum of $a(x,\omega)$ (red) 
versus that of the solution $p(x, \omega) = p(x,\bm\xi(\omega))$ (black);
we report the first 100 normalized eigenvalues with correlation length 
of $1/4$ for the log-parameter 
field $a(x, \omega)$. 
Bottom row: the left image shows 
the pointwise variance of $p(x, \bm\xi)$, computed using KL 
expansion of varying truncation levels, against 
pointwise variance computed using  $10^4$ Monte Carlo samples,
and the right image  
shows the pointwise variance of solution $p$ as we increase
the number of the finite-element grid points $N_d$.} 
\label{fig:pfield}
\end{figure*} 

Next, we study the properties of the PDE solution $p(x, \bm\xi)$.  We depict
the correlation function of $p(x,\bm\xi)$, approximated via Monte Carlo
sampling (with $10^4$ samples), in Figure~\ref{fig:pfield}~(top left).
This indicates strong correlations in the output field $p(x, \bm\xi)$.
%
%
%
In Figure~\ref{fig:pfield}~(top right) we compare
the (normalized) eigenvalues of the covariance operators for $p(x, \bm\xi)$ and 
$a(x, \omega)$; we note a much faster spectral decay for the output covariance
operator.
The latter indicates that a KL expansion with a small number of 
terms can be used to approximate $p(x, \bm\xi)$ reasonably well. We study this 
by considering 
the KL expansion 
\begin{equation} \label{kl_of_u}
    p(x, \bm\xi) = \bar{p}(x) + \sum_{j = 1}^\infty \sqrt{\lambda_j(C_p)} p_j(\bm\xi) v_j(x)
\end{equation}
of $p(x,\bm\xi)$,
where  $(\lambda_j(C_p), v_j)$ are the eigenpairs of covariance operator $C_p$ of $p$,
computed numerically using Algorithm~\ref{alg:KLE}, $p_j$ are given by 
\[
    p_j(\bm\xi) = \frac{1}{\sqrt{\lambda_j(C_p)}} \int_D (p(x, \bm\xi) - \bar{p}(x)) v_j(x) \, dx,
\quad j = 1, 2, \ldots,
\]
and $\bar{p}(x)$ is the mean of $p(x, \bm\xi)$.  

To quantify the impact of
truncating the KL expansion of $p(x, \bm\xi)$ on its approximation properties,
we study the pointwise variance $\var[p(x, \bm\xi)]$ with different choices of
$\Nklp$. Note also that \begin{equation}\label{eq:point_var} \var\left[ \sum_{j
= 1}^\Nklp \sqrt{\lambda_j(C_p)} p_j(\bm\xi) v_j(x)\right] = \sum_{j=1}^\Nklp \lambda_j(C_p)
v_j(x)^2.  
\end{equation} 
The results in Figure~\ref{fig:pfield}~(bottom left), indicate that pointwise variance of $p(x, \bm\xi)$ can be approximated well with a small $\Nklp$.  
Note that the finite-element grid resolution used to solve the PDE also affects
the accuracy the KL
expansion of the output. In Figure~\ref{fig:pfield}~(bottom right) we perform a
grid refinement study as we compute the pointwise variance of the process,
where we fix $\Nklp = \Nkla = 10$.  For the present problem using about 50 grid
points seems to be sufficient to resolve the pointwise variance.  More broadly,
one needs a sufficiently fine computational grid to ensure  the dominant
eigenpairs of the covariance operator are resolved with sufficient accuracy.
The grid resolution issues become more consequential in problems
in two or three space dimensions, as the dimension of the discretized
eigenvalue problem can become quite large.

\begin{figure*}[ht]\centering
\includegraphics[width = .3\textwidth]{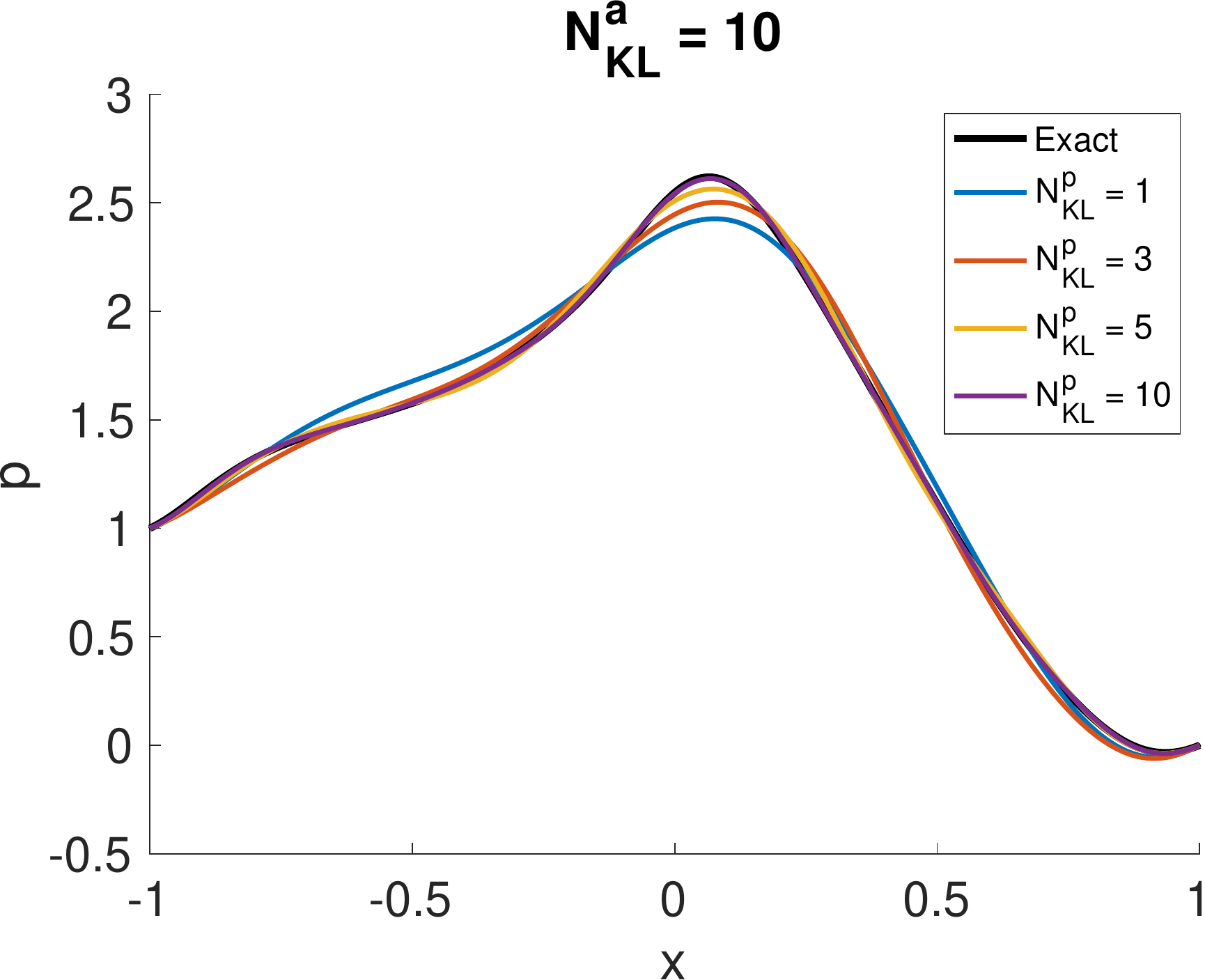}	
\includegraphics[width = .3\textwidth]{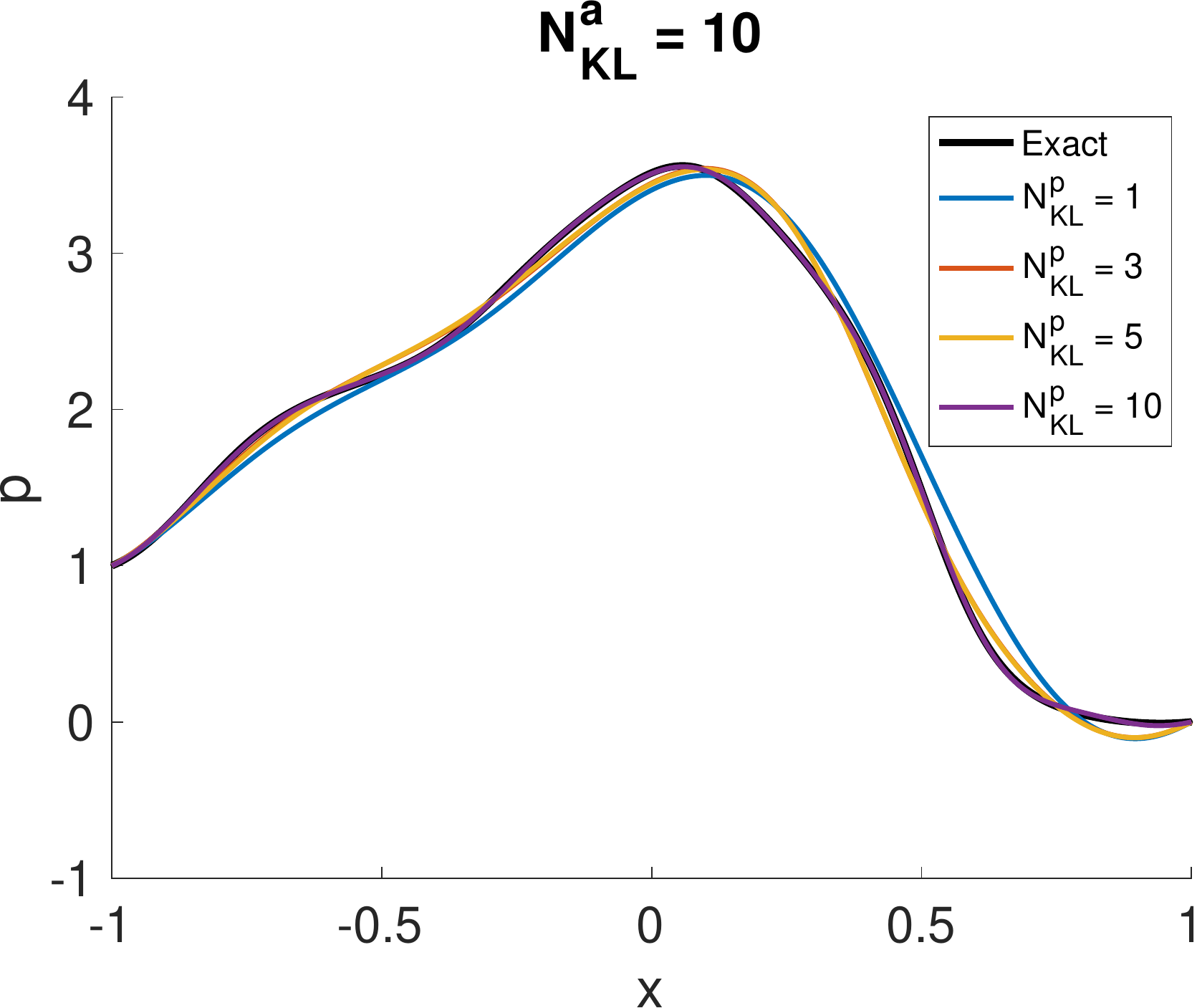}	
\includegraphics[width = .3\textwidth]{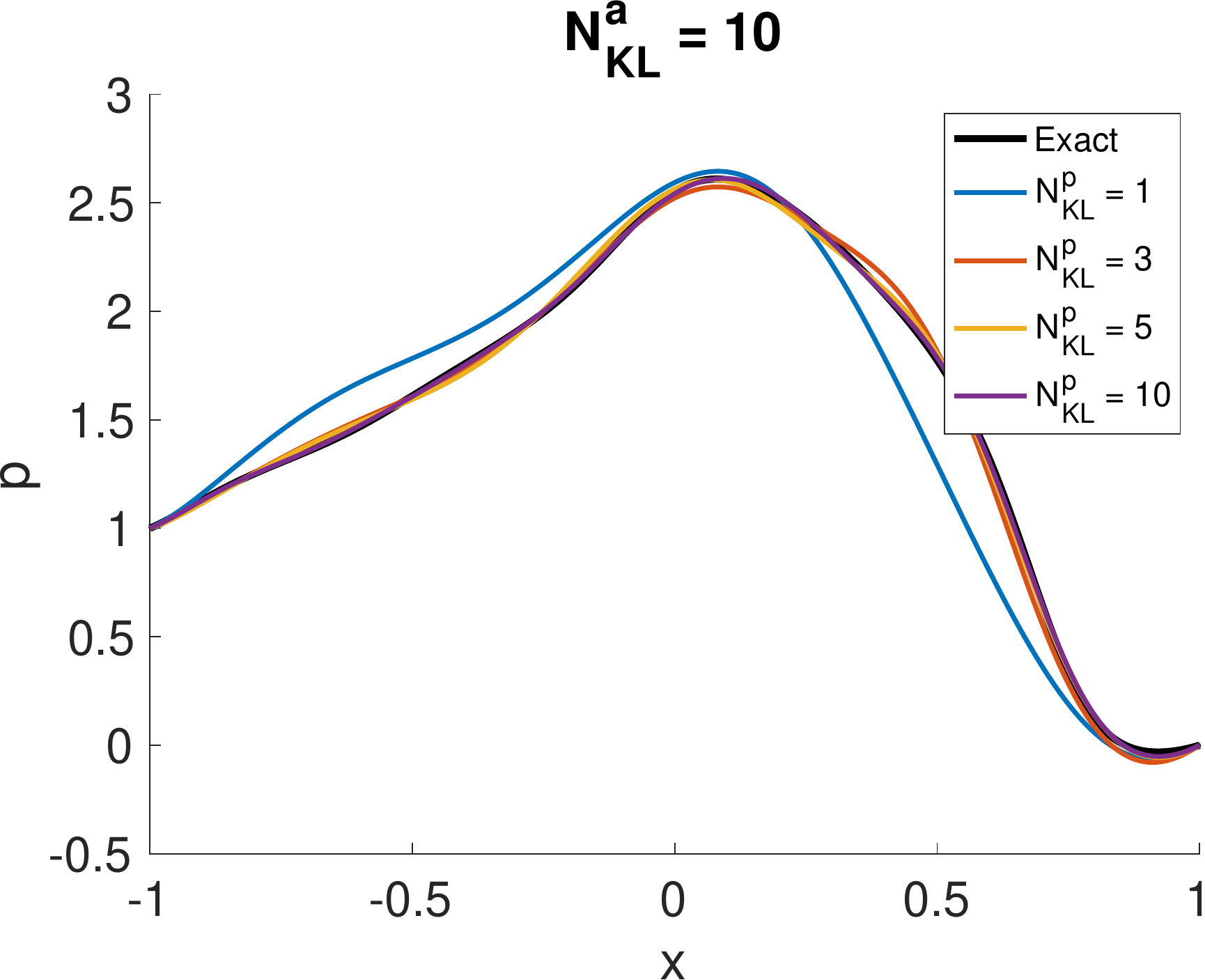}\\	
\includegraphics[width = .3\textwidth]{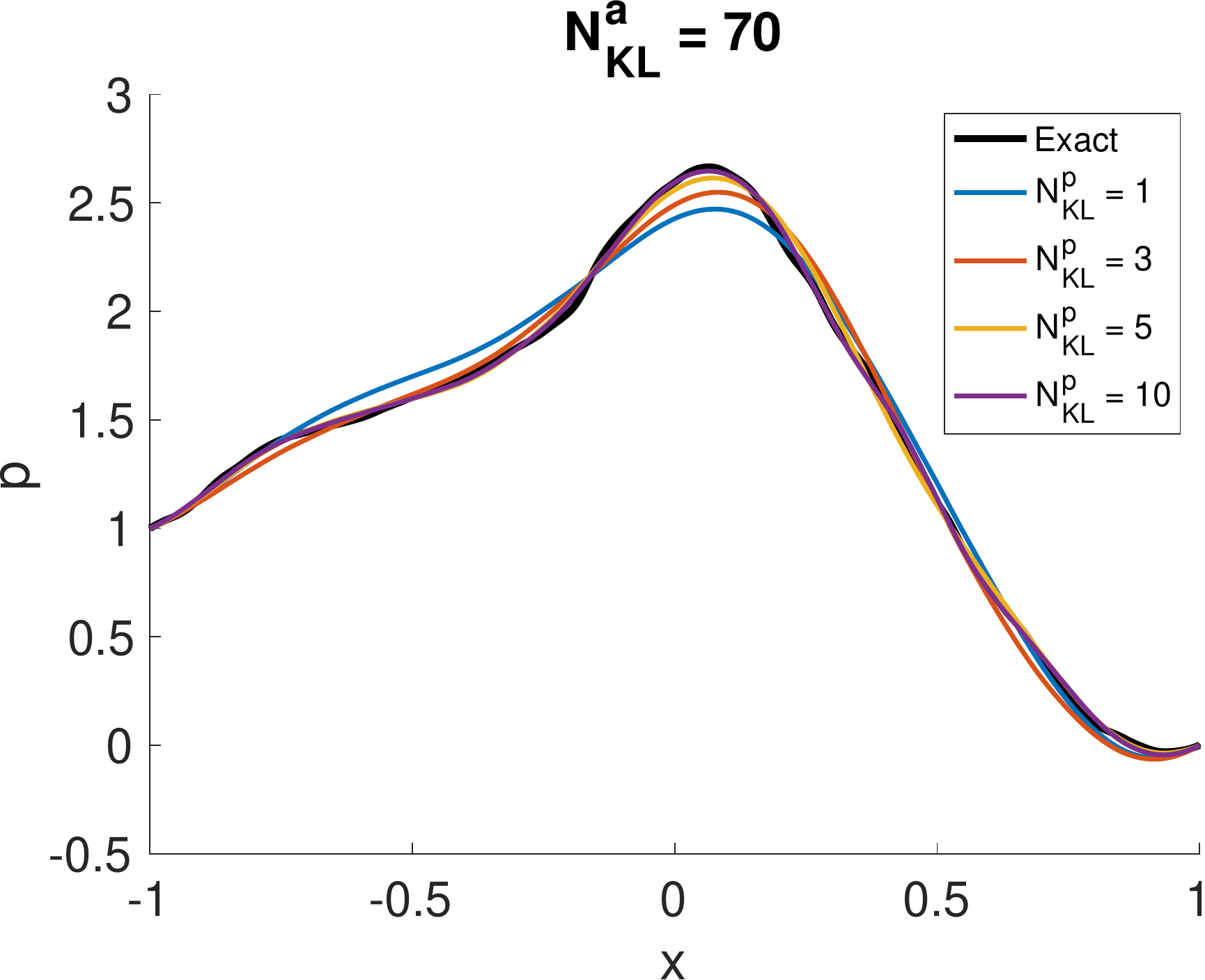}	
\includegraphics[width = .3\textwidth]{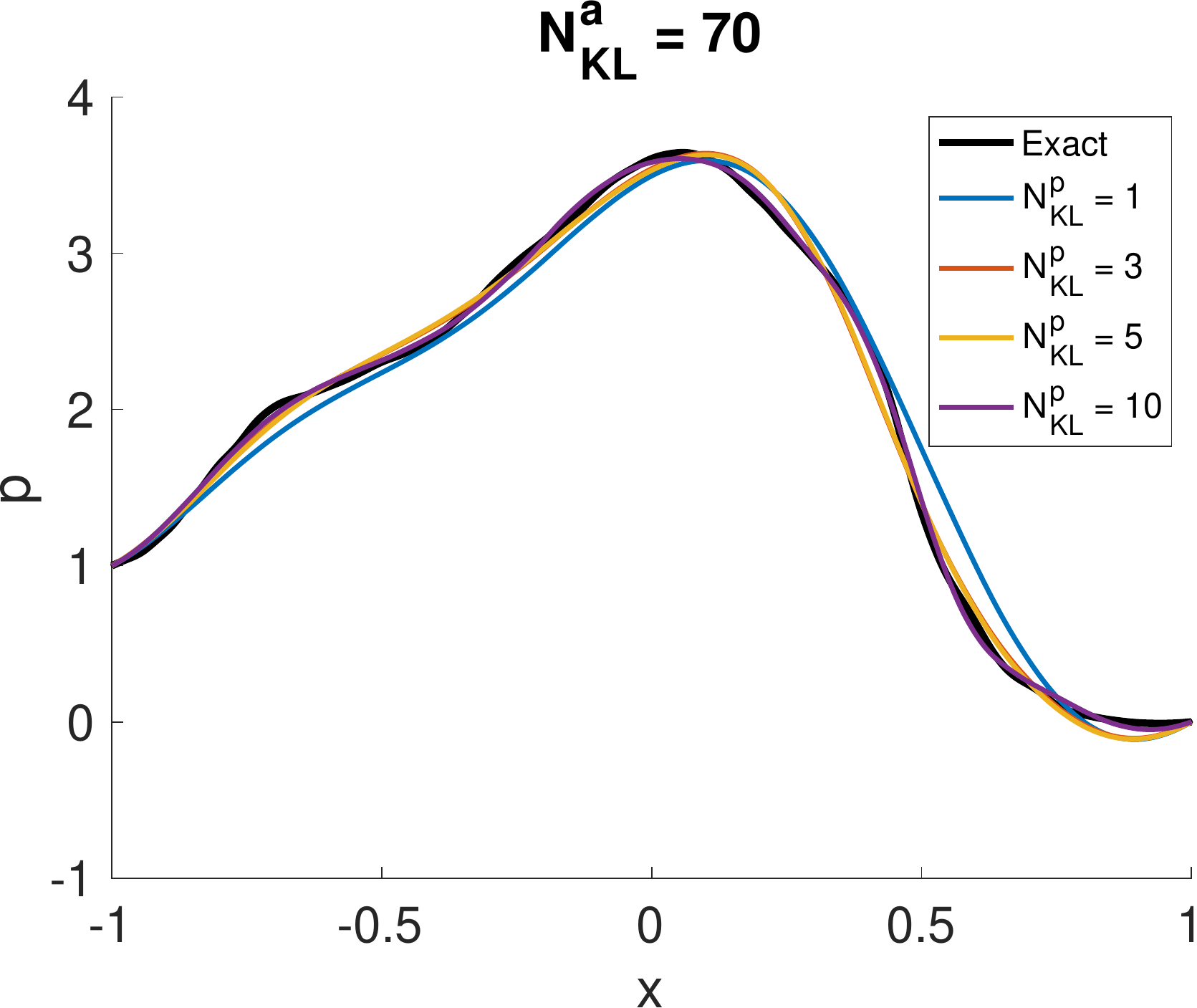}	
\includegraphics[width = .3\textwidth]{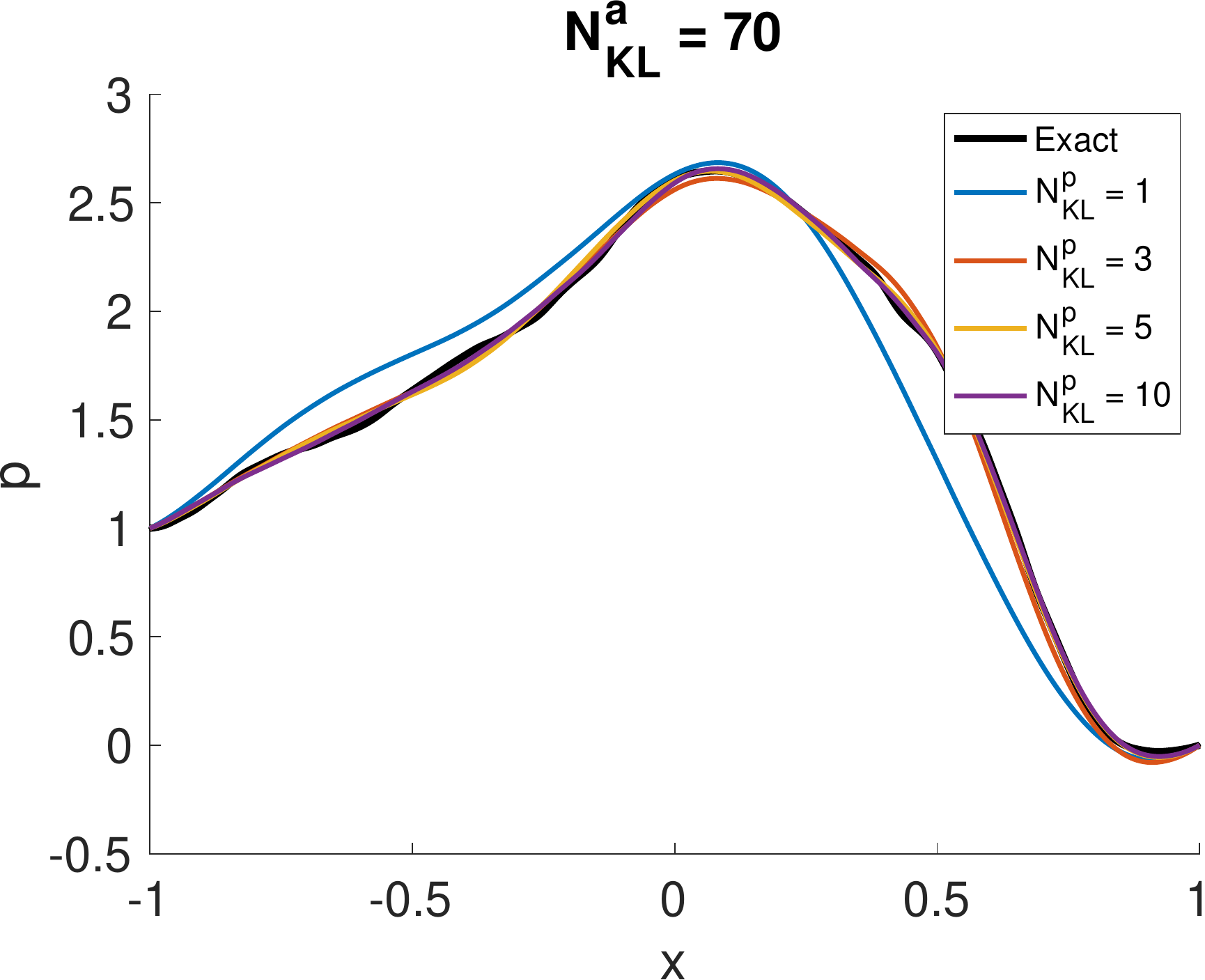}	
\caption{Three realizations of $p(x, \bm\xi)$ and the corresponding truncated 
KL expansions of $p$; each row corresponds to approximations computed with 
different truncation levels for the parameter, as indicated by $N_{KL}^a$
in figure titles.}
~\label{fig:paramoutput_reduction}
\end{figure*}

We next study input parameter and output
dimension reduction in Figure~\ref{fig:paramoutput_reduction} where we show
typical realizations of $p(x, \bm\xi)$, for a small $\Nkla$ (top row) 
and a relatively large $\Nkla$ (bottom row) for various choices of 
$\Nklp$ (output dimension). 

The numerical experiments in this section lead to the following observations:
(i) it is possible to reduce parameter dimension by focusing on KL terms of the
parameter that the PDE solution operator is most sensitive to; and (ii) it is
possible to reduce output dimension by focusing on the dominant KL terms of the
output.  In the next section, we explore these notions systematically, 
in a more challenging
problem, involving biotransport in tumors.

\section{APPLICATION TO BIOTRANSPORT IN TUMORS}
\label{sec:biotransport}

\textbf{Governing equations and numerical setup.} In this section, we study the pressure field in a tumor when a single needle
injection occurs at the tumor center. A 2D model in a polar
coordinate system is used to analyze the flow field. Consider the mass
conservation law and Darcy's law for steady incompressible flows in a 2D
domain, $D = \{(r, \theta) : R_\text{needle} < r < R_\text{tumor},\,
0 < \theta < 2\pi\}$,
\begin{equation}\label{equ:2D_Darcy}
\frac{\partial}{\partial r} \left(\frac{\kappa r}{\mu} \frac{\partial p}{\partial r}\right) +
   \frac{1}{r} \frac{\partial}{\partial \theta} \left(\frac{\kappa }{\mu} \frac{\partial p}{\partial \theta}\right)
   =0. 
\end{equation}
Here $p$ is the pressure, $\kappa$ is the permeability, $\mu$ is the fluid dynamic
viscosity, $r$ is the radial distance from a fixed origin, $\theta$ is the polar angle, 
$R_\text{tumor}$ is the radius of the tumor, and $R_\text{needle}$ is the radius
of the needle used to inject nanofluid into the tumor.  

The boundary conditions for the pressure equation are specified as follows: 
\begin{equation}\label{equ:DarcyBC}
\left\{
\begin{aligned}
     p&=0, &\quad r=R_\text{tumor},\\
     \frac{\partial p}{\partial r}&= \frac{-Q \mu}{2 \pi R_\text{needle} \kappa}, &\quad r=R_\text{needle}.
\end{aligned}
\right.
\end{equation}
Herein, $Q$ is the volume flow rate per unit length.  Periodic boundary
conditions are enforced in the $\theta$ direction. In this study,
$R_\text{needle}$ and $R_\text{tumor}$ are set to $0.25~mm$ and $5~mm$,
respectively, $Q$ is $1~mm^2/min$, 
and $\mu$ is $8.9 \times 10^{-4}~Pa \cdot s$.

\textbf{Uncertainties in permeability field.}
As before, let $(\Omega, \mathcal{F}, P)$ be a probability space. 
Following~\cite{AlexanderianZhuSalloumEtAl17}, the permeability field $\kappa$ is 
modeled by a log-Gaussian random field, and its mode is set to $0.5~md$, 
where $md$ stands for millidarcy.
We assume that the log-permeability, 
$a(\vec{x}, \omega) = \log\big(\kappa(\vec{x}, \omega))$,
is given by 
\[
a(\vec{x}, \omega) = a_0(\vec{x}) + \sigma_a Z(\vec{x}, \omega),
\quad \vec{x} \in D, \omega \in \Omega.
\]
Here $a_0$ is the pointwise mean of the process, $\sigma^2_a$ is the pointwise
variance, and $Z$ is a centered Gaussian process with unit pointwise variance
for every $\vec{x} \in D$.  In this study, $\sigma_a^2$ is  set to $0.25$, and $a_0$
is calculated from the definition of the mode of $\kappa$ as $a_0 = \ln(0.5) +
\sigma_a^2$.  The covariance function of $Z$ is expressed as $c_Z(\vec{x},
\vec{y}) = \exp\left\{- \frac{1}{\ell} \| \vec{x} - \vec{y}
\|_1\right\}$, $\vec{x}, \vec{y} \in D$, where $\ell > 0$ is the correlation
length.  As before, the (Gaussian) log-permeability field can be expressed with a truncated KL
expansion,
\begin{equation}\label{equ:2DKLE}
    a(\vec{x}, \omega) \approx a_0(\vec{x}) + \sigma_a \sum_{i = 1}^\Nkla \sqrt\lambda_i 
                                              \xi_i(\omega) e_i(\vec{x}),
\end{equation}
where $\lambda_i$ and $e_i$ are eigenpairs of the covariance operator of 
$Z$, and $\xi_i$ are independent
standard normal random variables. 

\begin{figure*}[ht]\centering
\includegraphics[width=.3\textwidth]{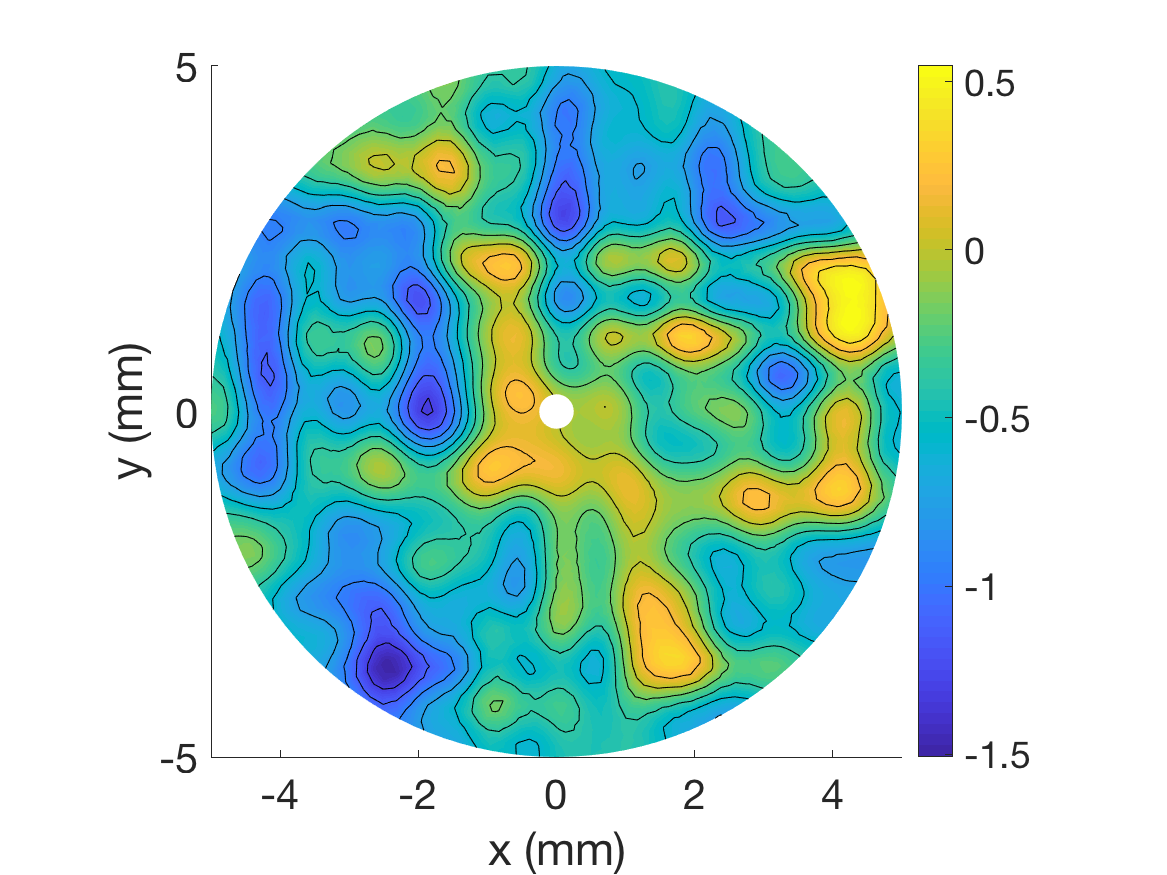}
\includegraphics[width=.3\textwidth]{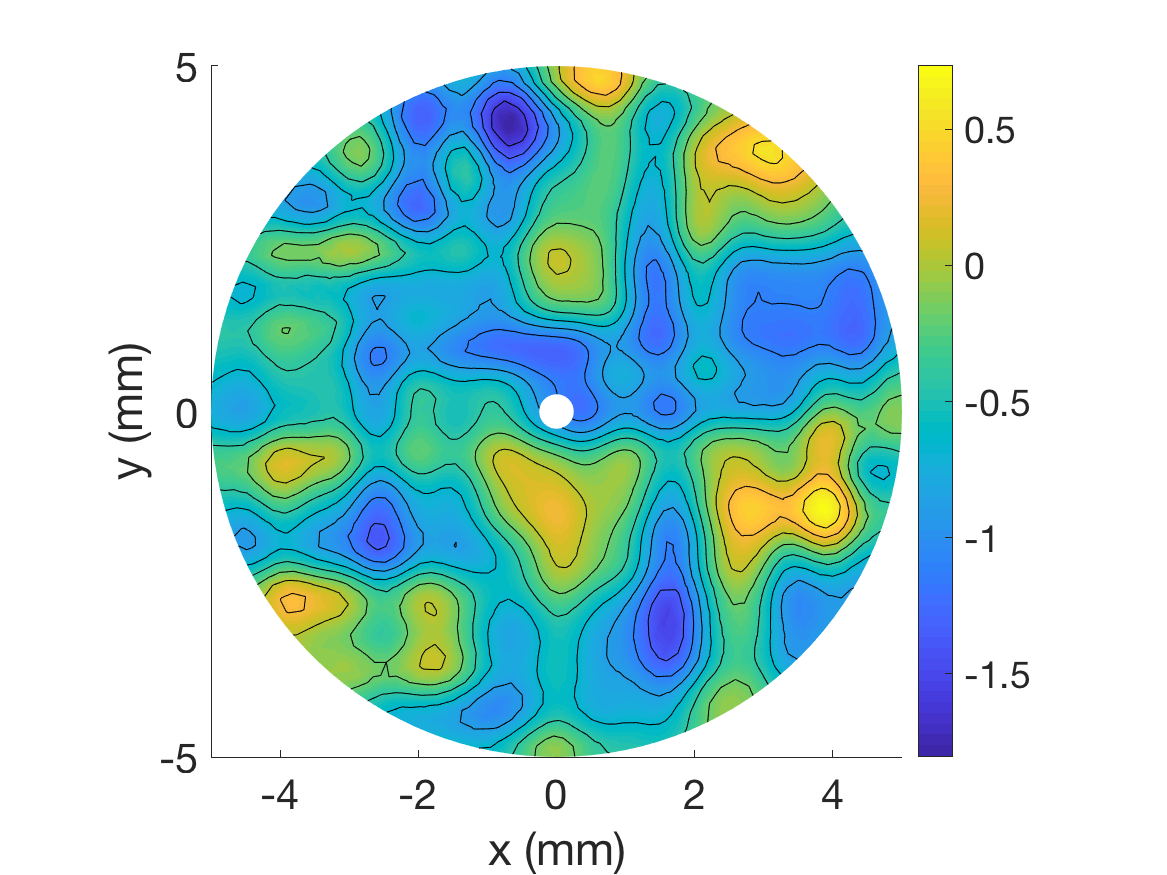}
\includegraphics[width=.3\textwidth]{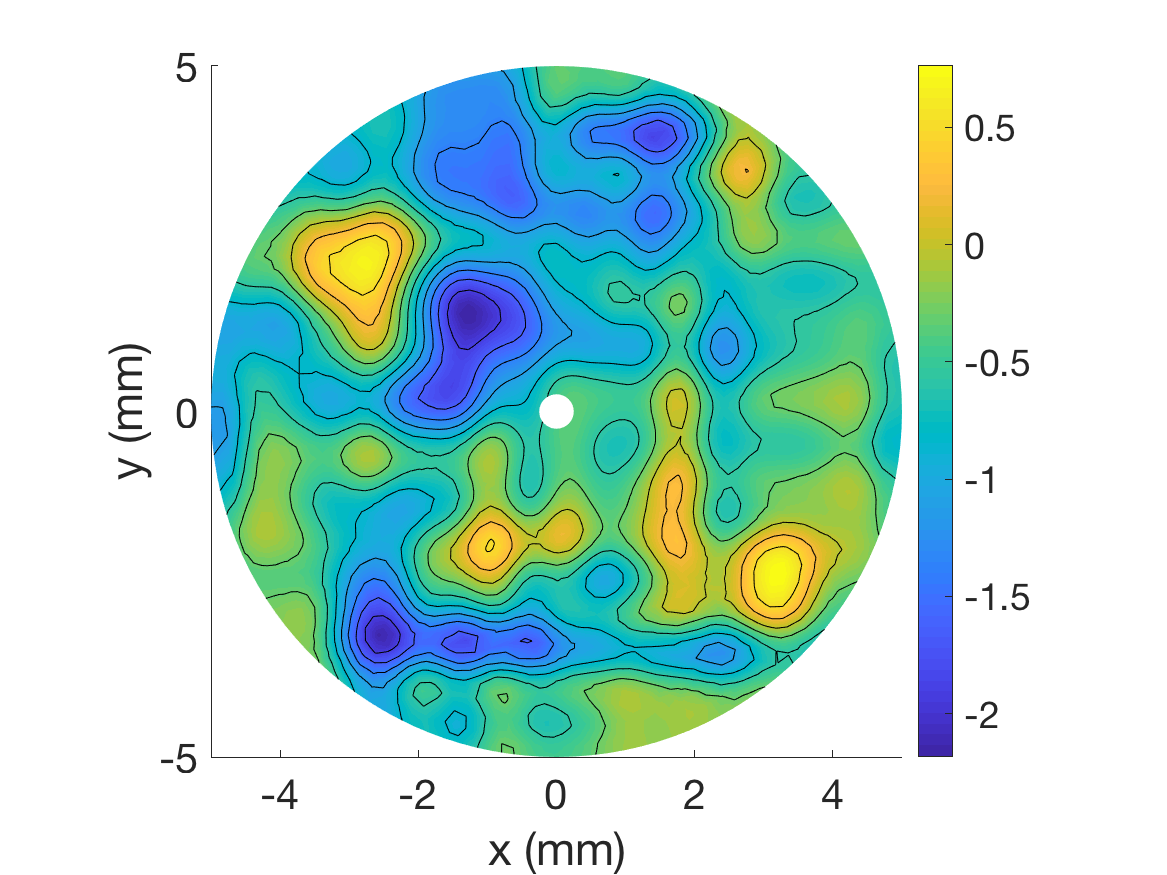}\\
\includegraphics[width=.3\textwidth]{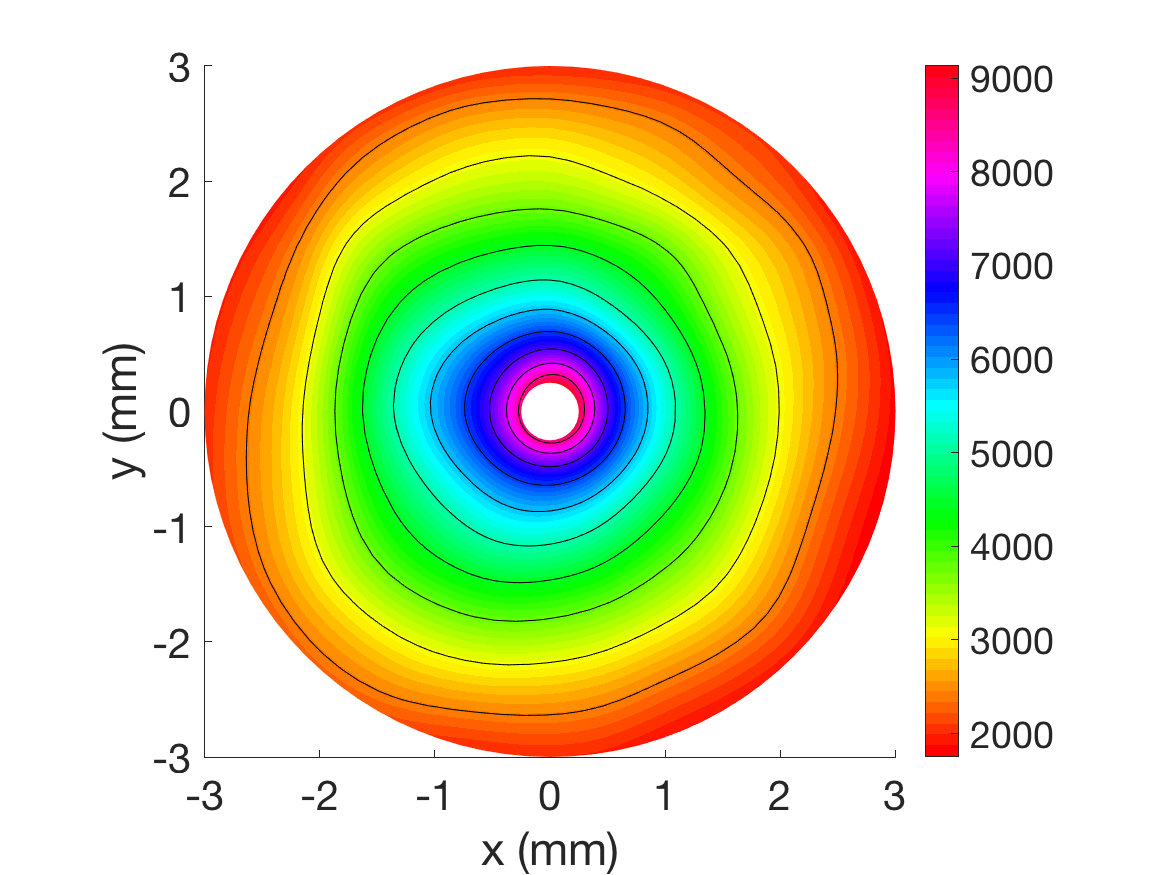}
\includegraphics[width=.3\textwidth]{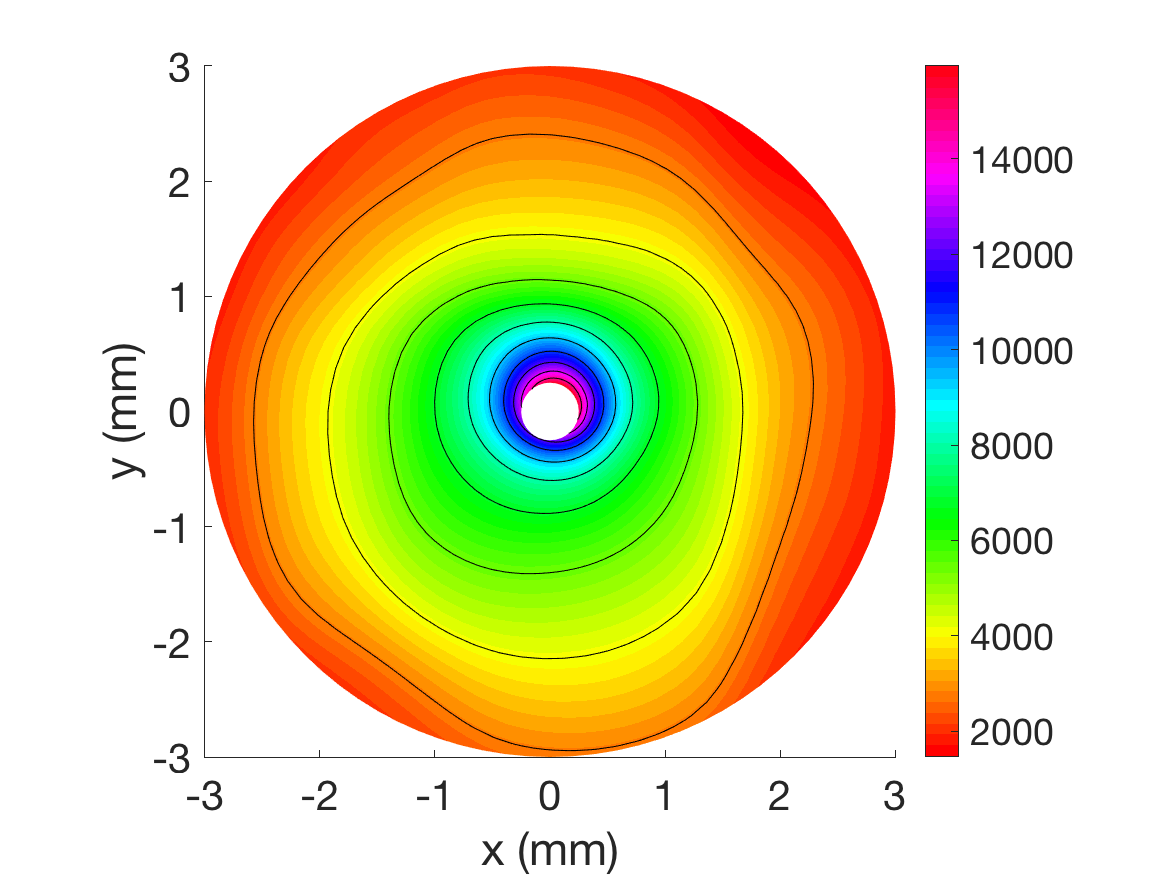}
\includegraphics[width=.3\textwidth]{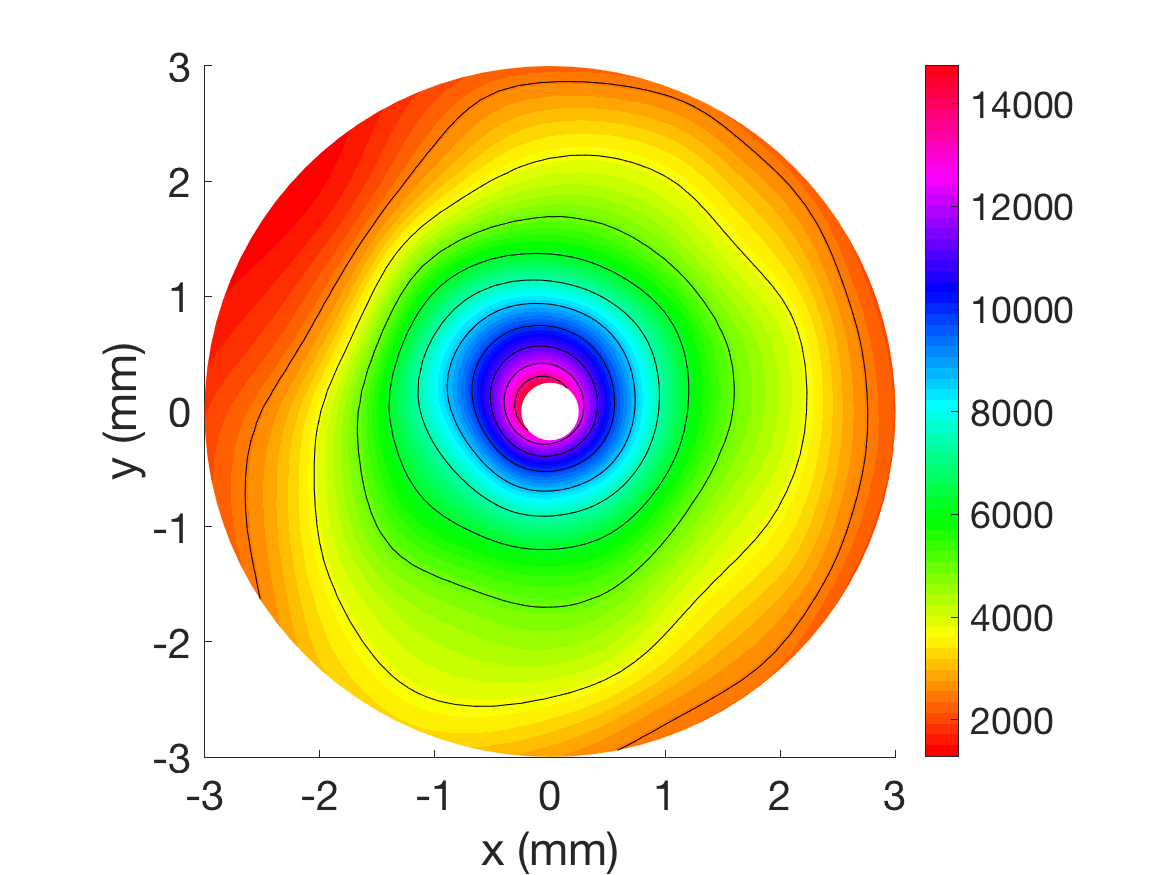}
\caption{Three sets of realizations of the permeability field (top) and the corresponding pressure 
field restricted to a subdomin with outer radius of $R_{qoi} = 3$ mm (bottom). 
The correlation length $\ell$ is $1$ mm.}
\label{fig:2Drealizations}
\end{figure*}

\textbf{Uncertainty properties of the pressure field.} Whereas the governing equation is more complex than the previously considered
1D problem, it is still an elliptic PDE and hence we observe similar
behavior in terms of potential for dimension reduction. In the present
example, we focus on pressure field over regions around the injection 
site. Specifically, we consider annular regions with inner boundary given by
the inner boundary of the domain and the outer boundary specified
by circles of radius $\Rq = 1$ mm, $\Rq = 2$ mm, or $\Rq = 3$ mm.
Three sets of
realizations of the permeability field (in the entire domain) and the corresponding model output 
(in the annular domain with $\Rq = 3$ mm) 
are presented 
in Figure~\ref{fig:2Drealizations}. 
We observe that
although the permeability field realizations exhibit complicated features, the
fluctuations in the pressure field are mild. 

\begin{figure}[ht]\centering
\includegraphics[width=0.32\textwidth]{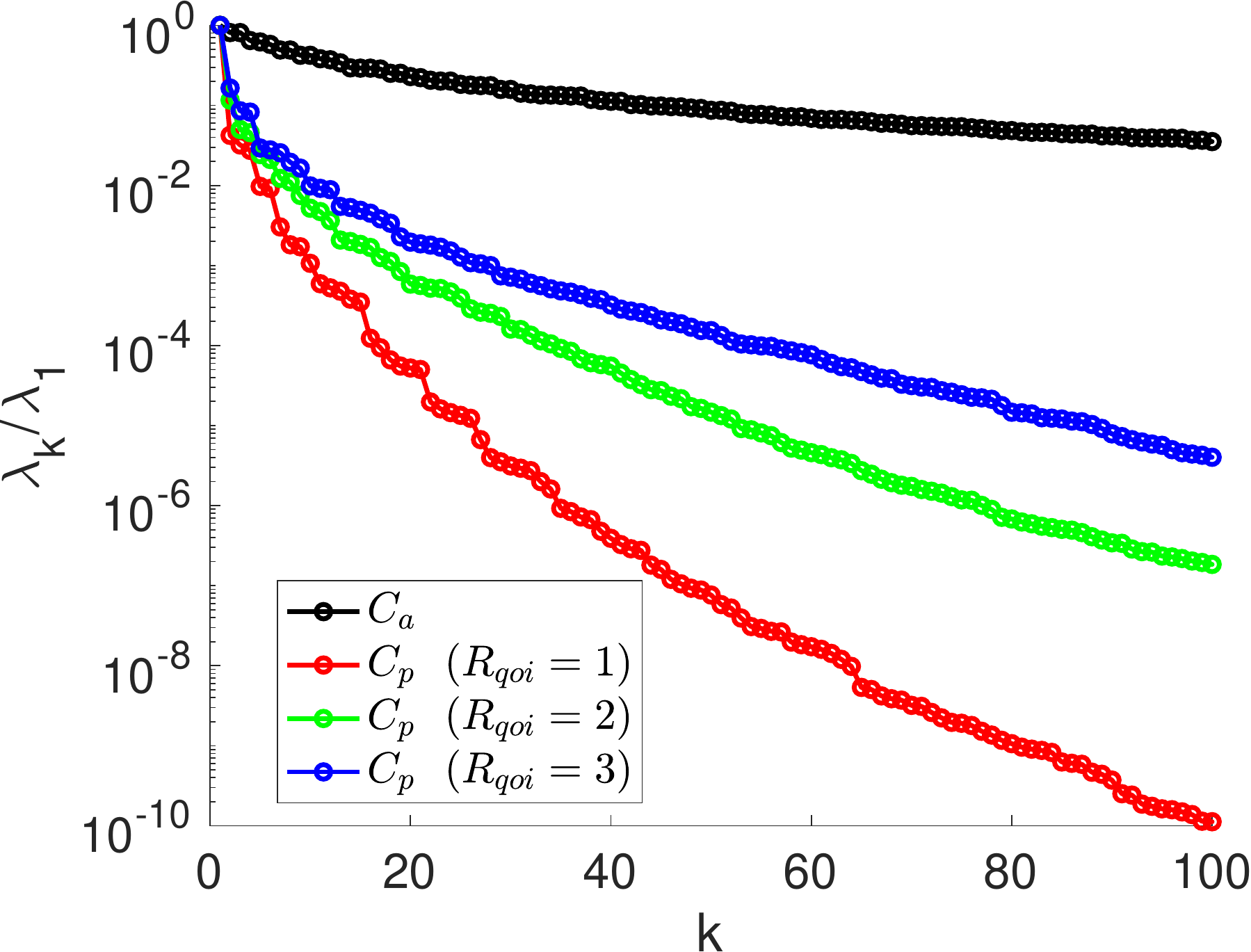}
\includegraphics[width=0.32\textwidth]{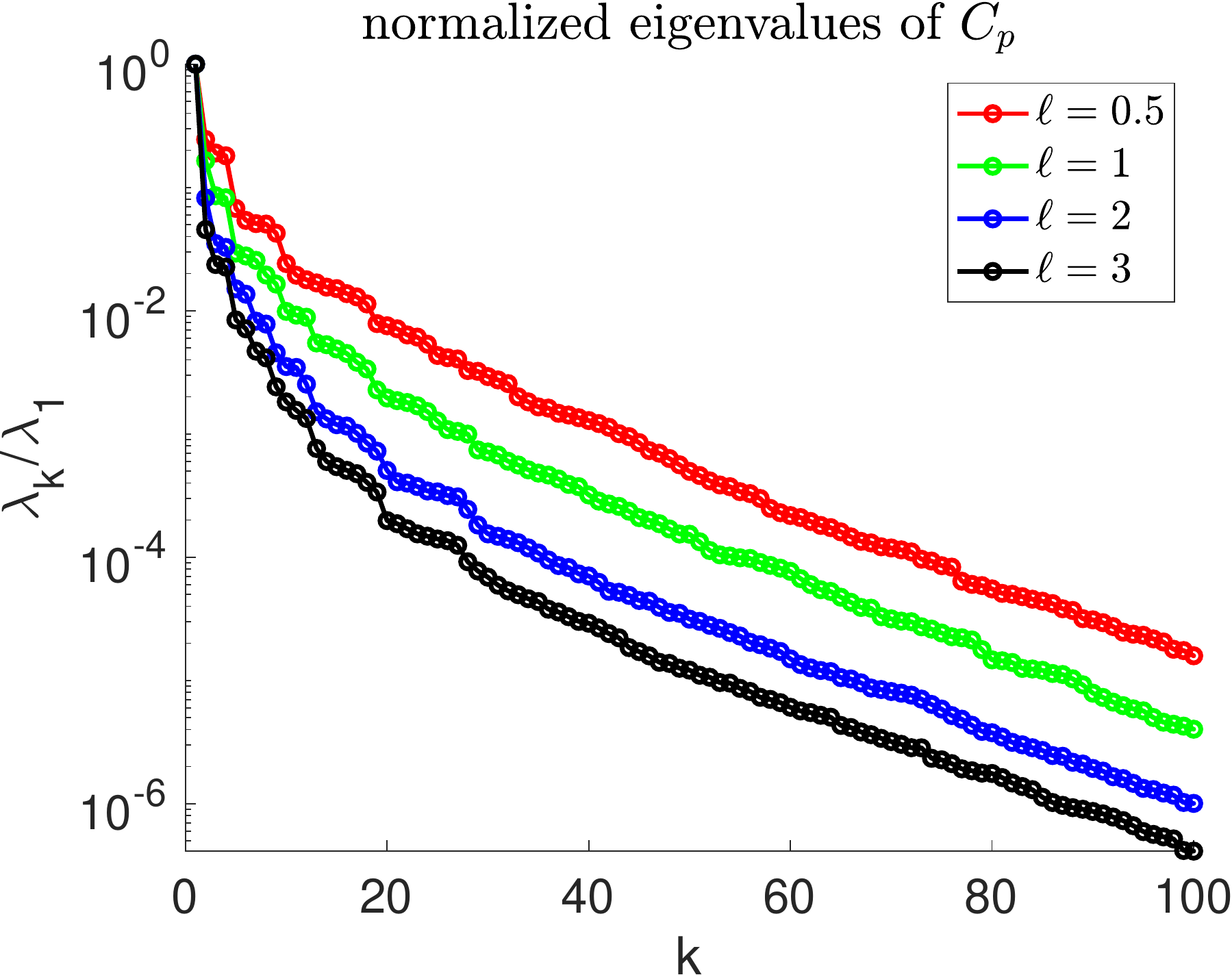}
\includegraphics[width=0.32\textwidth]{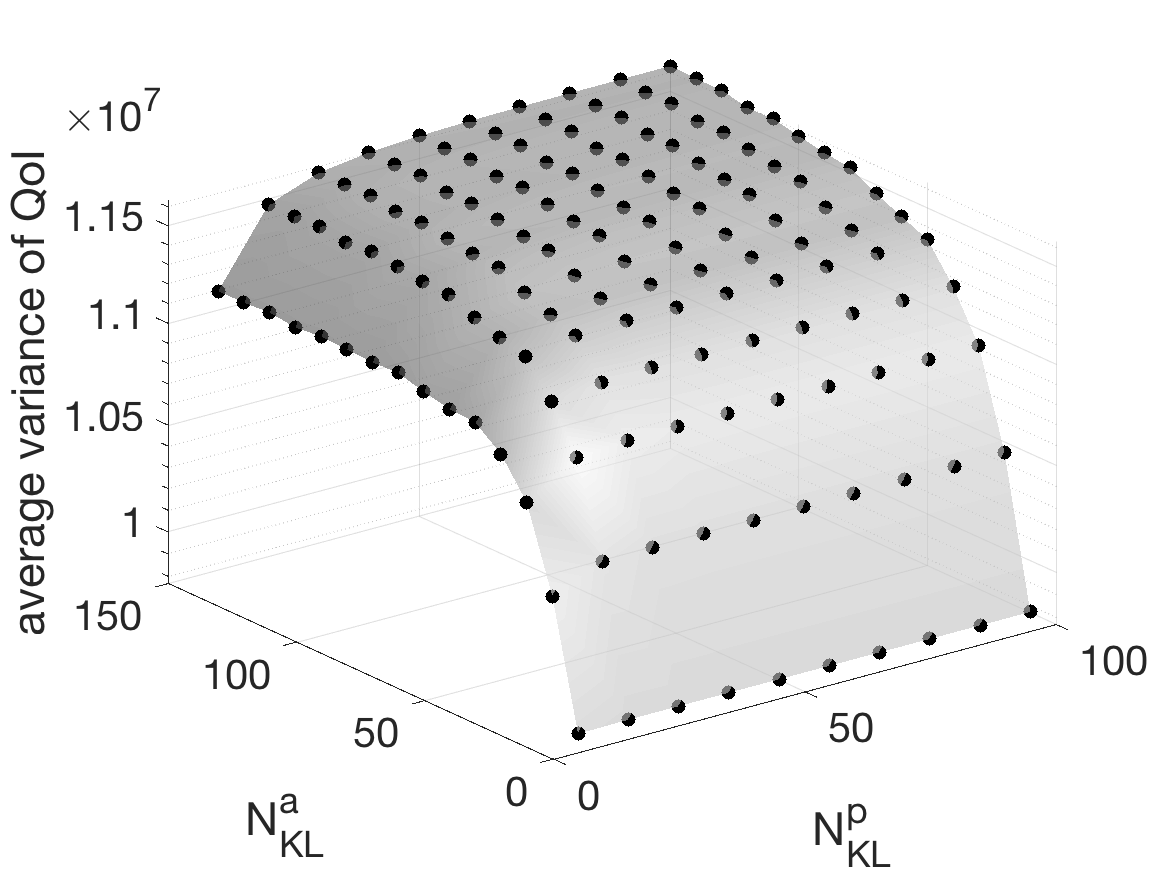}
\caption{Top:  
spectrum of $C_a$ versus
that of $C_p$, corresponding to regions of interest with different sizes;
middle: spectrum of $C_p$ corresponding to different input parameter correlation
lengths; bottom: 
the average variance of the output, captured by its truncated 
KL expansion, as we increase
$\Nkla$ and $\Nklp$.} 
\label{fig:UQ_2d}
\end{figure}

We denote the covariance operator of the log-permeability field by $C_a$ and
that of the pressure field by $C_p$.  In Figure~\ref{fig:UQ_2d}~(top), we
report the (normalized) eigenvalues of $C_a$ and those of $C_p$, corresponding
to $\Rq = 1$ mm, $\Rq = 2$ mm, and $\Rq = 3$ mm. First, we note that the
eigenvalues of $C_p$ exhibit a far more rapid decay as compared to that of
$C_a$.  Moreover, as the size of region of interest decreases, the spectral
decay of $C_p$ becomes faster. 
In Figure~\ref{fig:UQ_2d}~(middle), we examine the spectral decay of $C_p$, as
the correlation length of the log-permeability increases; for this test we used
$\Rq=3$ mm.  As expected, increasing the complexity of the input parameter, by
using smaller correlation lengths, leads to slower spectral decay for $C_p$.
However, we find that even with $\ell = 0.5$ mm, about $96\%$ of average output
variance is captured by the first 20 KL modes of the output. 
Finally, Figure~\ref{fig:UQ_2d}~(bottom) summarizes  the effect of input and
output dimensions on capturing the average variance of the output (with $\Rq =
3$ mm, and input parameter correlation length $\ell = 1$ mm).  Note that the
variance of $p(\vec{x}, \bm\xi)$, restricted to the region of interest, is
computed by $\sum_{k = 1}^\Nklp \lambda_k(C_p)$.  We note that the average
variance can be approximated with reasonable accuracy with small $\Nkla$ and
$\Nklp$.

\begin{figure}[ht]\centering
\includegraphics[width=0.32\textwidth]{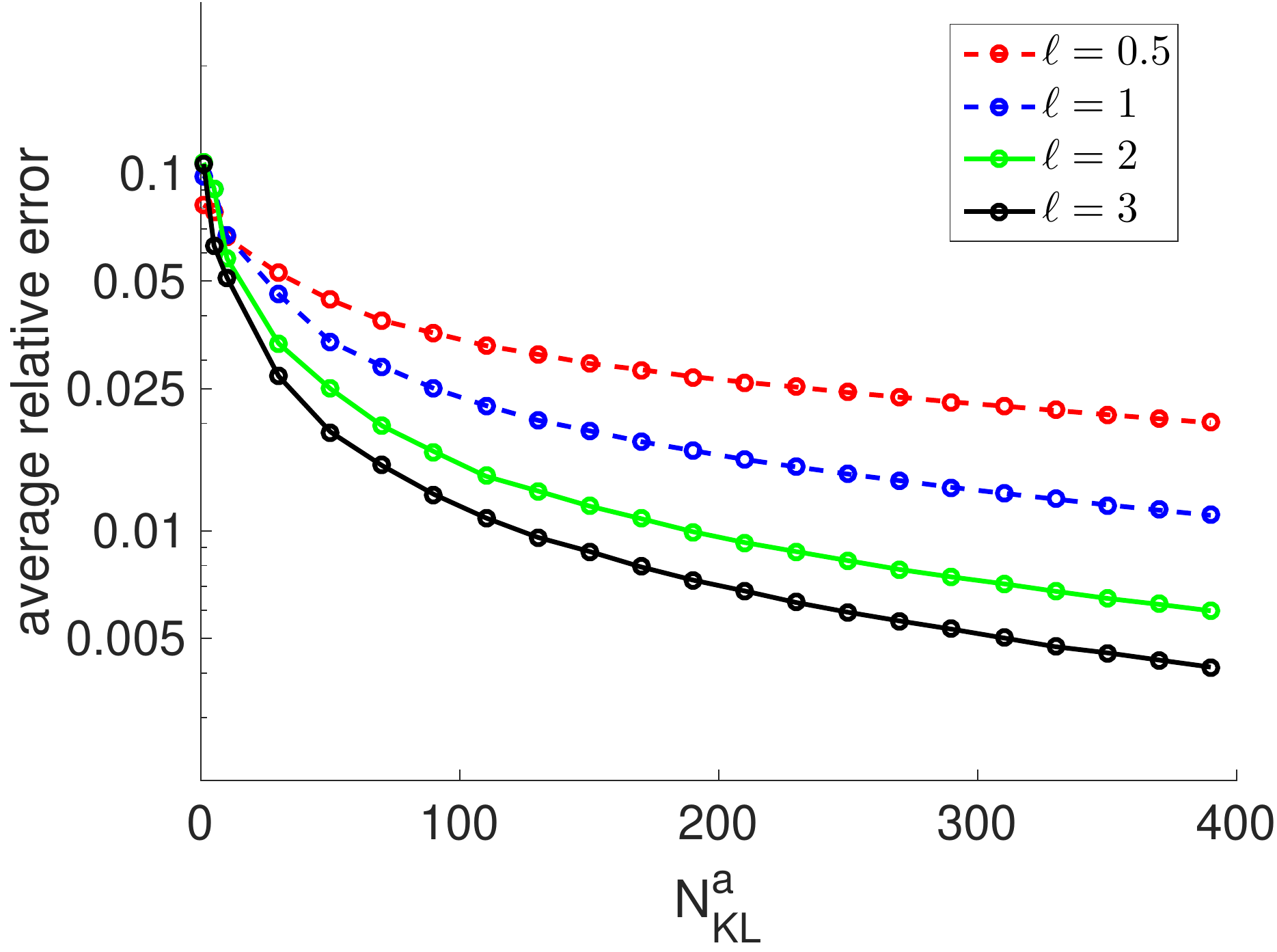}
\includegraphics[width=0.32\textwidth]{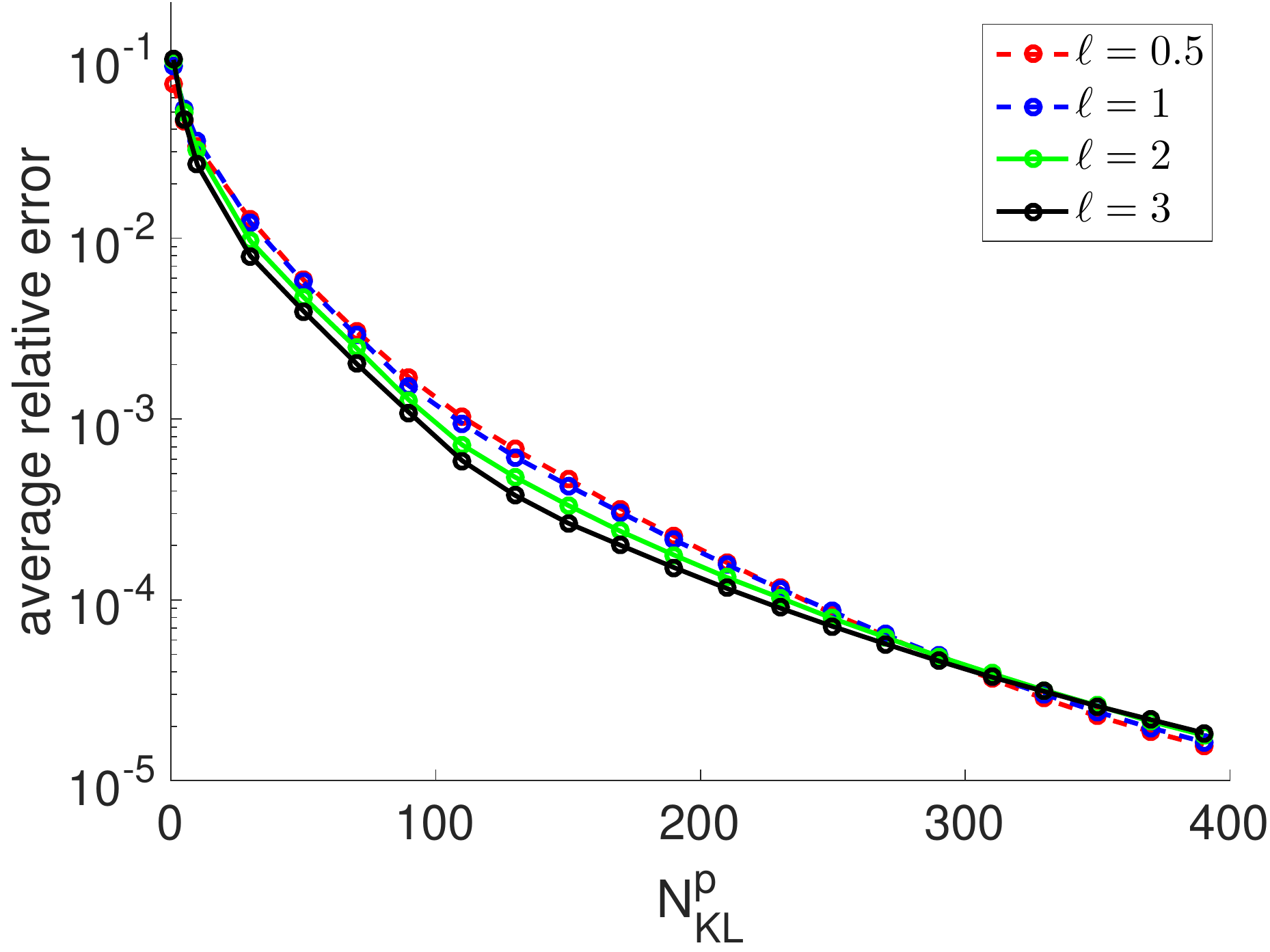}
\caption{Accuracy of the KL representation of the output (with $\Rq = 3$ mm)
as $\Nkla$ (top) and $\Nklp$ (bottom) increase. The average $L^2$ errors
were computed using $1{,}000$ Monte Carlo samples.}
\label{fig:approx_2d}
\end{figure}

We also examine the average relative $L^2$ error of the truncated KL
representation of the output (with $\Rq = 3$ mm) as $\Nkla$ and $\Nklp$ increase,
for input (i.e., permeability) fields with different correlation lengths; the results are reported in top
and bottom panels of Figure~\ref{fig:approx_2d}, respectively. For the
figure in top, we used the KL expansion of input with $2{,}000$ terms as a reference 
true log-permeability field. For the figure at the bottom, we compute the relative
$L^2$ error of the output KL representation with the PDE solution restricted to
the region of interest. We note that when the input dimension is fixed, 
the average relative error of the output KL expansion decreases very fast when the number of output KL modes increase, 
and is not very sensitive to input parameter correlation length. On the
other hand, for small correlation lengths, there is a notable increase in
the number of input KL modes needed to represent the output accurately. 


\textbf{Insights into reduced-order modeling.}
From previous analysis, we observed that the spectrum of the output covariance operator $C_p$ decays very fast, even when the correlation length $\ell$ is small. This indicates that the output, i.e., the pressure field, can be effectively approximated by a truncated KL expansion as
\begin{equation}\label{equ:KL_P}
    p(\vec{x}, \bm\xi) = \bar{p}(\vec{x}) + \sum_{j = 1}^\Nklp \sqrt{\lambda_j(C_p)} 
    p_j(\bm\xi) v_j(\vec{x}),
\end{equation}
with a small number $\Nklp$ of KL terms. The importance of this approximation
is that it decouples the spatial (i.e., $\vec{x}$) dimensions and those of the
random variable $\bm \xi (\omega)$. If a surrogate model, such as 
a polynomial chaos  expansion (PCE)~\cite{Ghanem,knio} 
is directly used to approximate the pressure field for the purpose of uncertainty
quantification~\cite{AlexanderianZhuSalloumEtAl17}, the PCE needs to be
built for each spatial point on the computational mesh. Instead, if the
approximation in \eqref{equ:KL_P} is used, PCE (or any other suitable
surrogate model) only needs to be constructed
for each KL mode $p_j$, $j=1,\ldots, \Nklp$.  

\begin{figure}\centering \includegraphics[trim=0 100 0
100,clip,width=.45\textwidth]{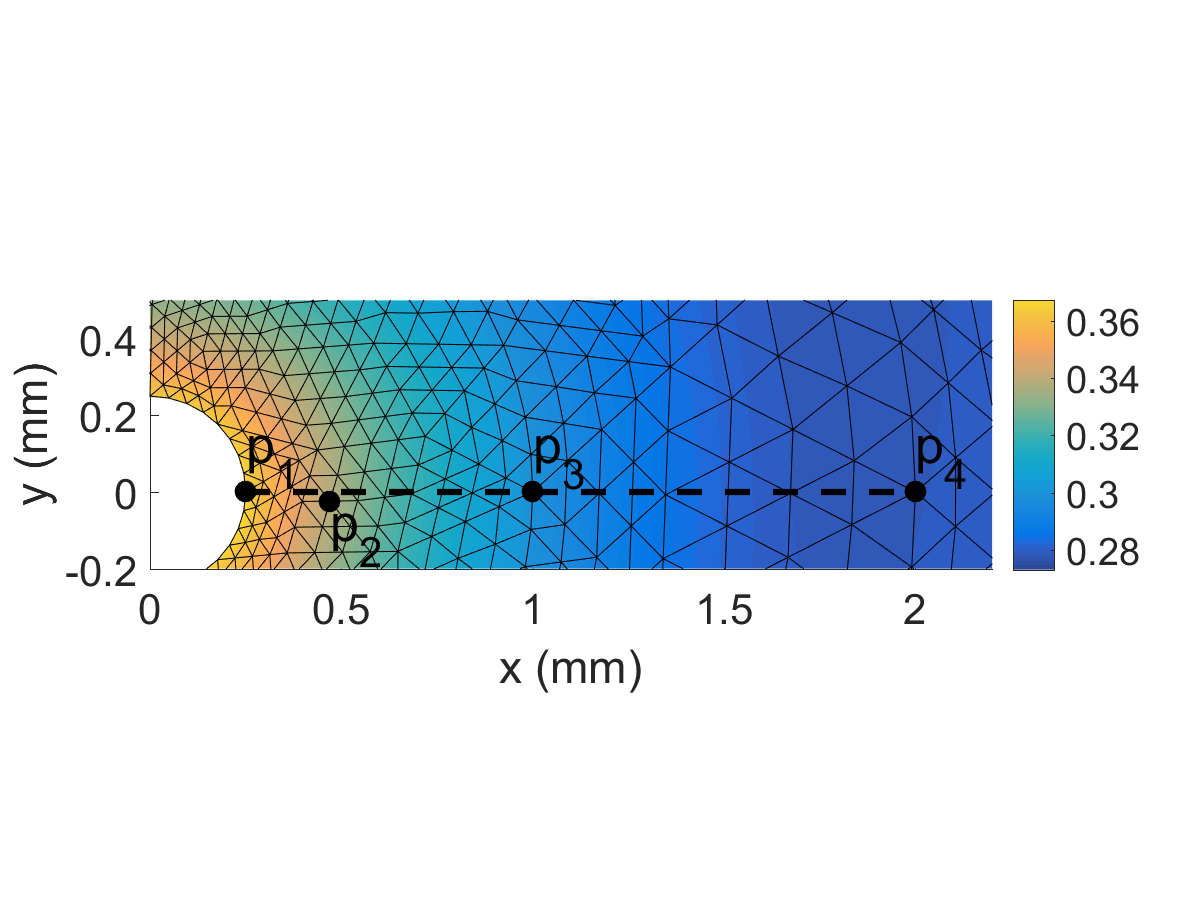}
\caption{Distribution of the points where PDFs of pressures are extracted, and
the corresponding RSD field (contour).} \label{fig:Points_Dis} \end{figure}

We evaluate the performance of the reduced-order model (ROM), i.e.,
\eqref{equ:KL_P}, on recovering the probability density functions (PDFs) of
pressures at different locations in the flow field.  The KL expansion of
output is computed using Algorithm~\ref{alg:KLE}. As shown 
in~\cite{AlexanderianReeseSmithEtAl18}, a modest sample size $N$ can be used
to capture the dominant modes of output KL expansion. Here, to ensure
accuracy, we use $N = 1{,}000$ samples. 
The input dimension is
fixed at $\Nkla = 150$ in following tests.  In Figure~\ref{fig:Points_Dis}, we
present four points, namely, $P_1$, $P_2$, $P_3$ and $P_4$, on the mesh where
PDFs are constructed. The contour stands for relative standard deviation (RSD)
of the pressure field. 
Two correlation lengths, namely, $\ell = 0.5$ mm and $\ell = 3$ mm,
are tested. In both cases, the region of the quantity of interest has a outer
radius $\Rq$ of $3$ mm. 

\begin{figure*}[ht]\centering
\includegraphics[width=.45\textwidth]{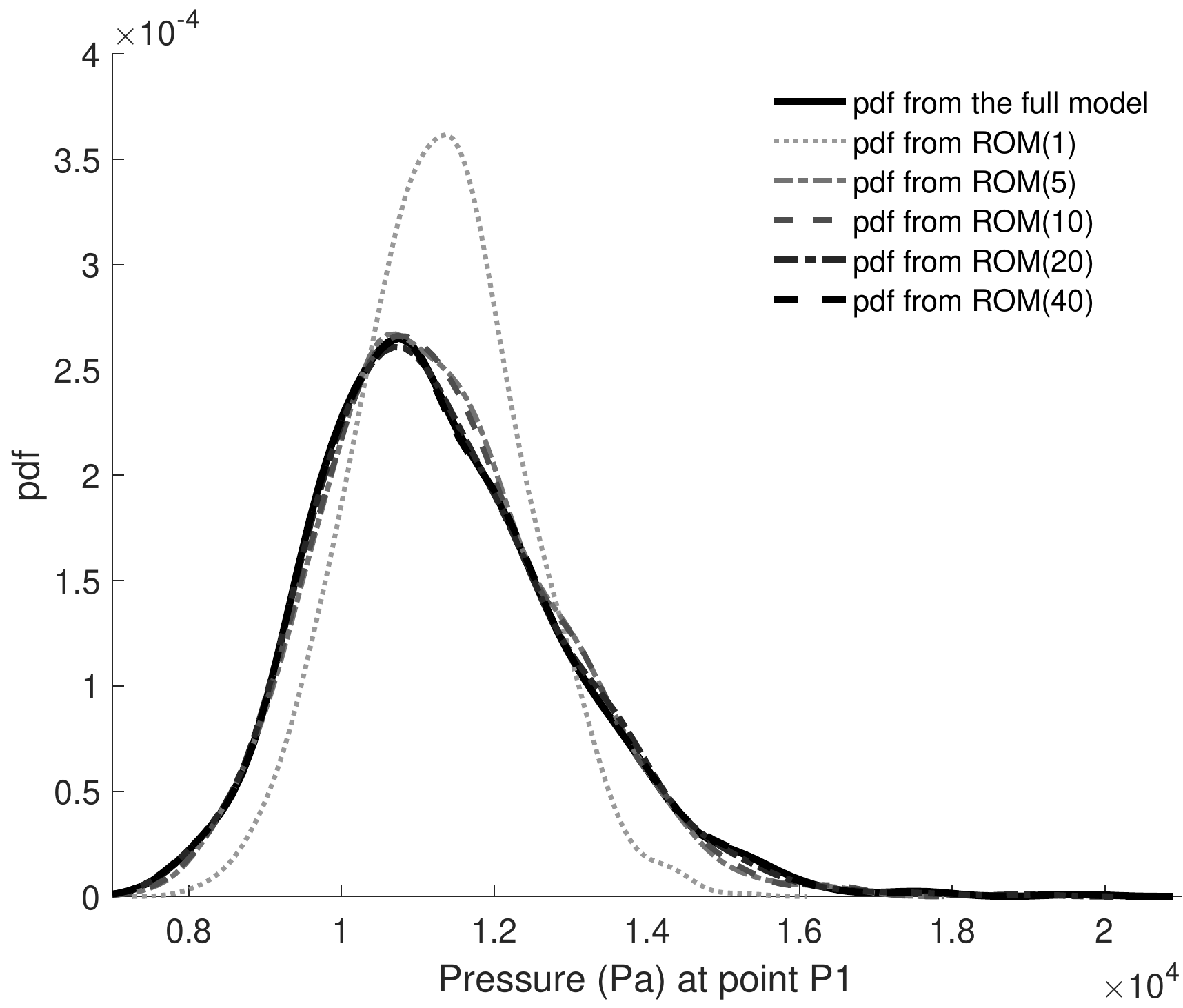}
\includegraphics[width=.45\textwidth]{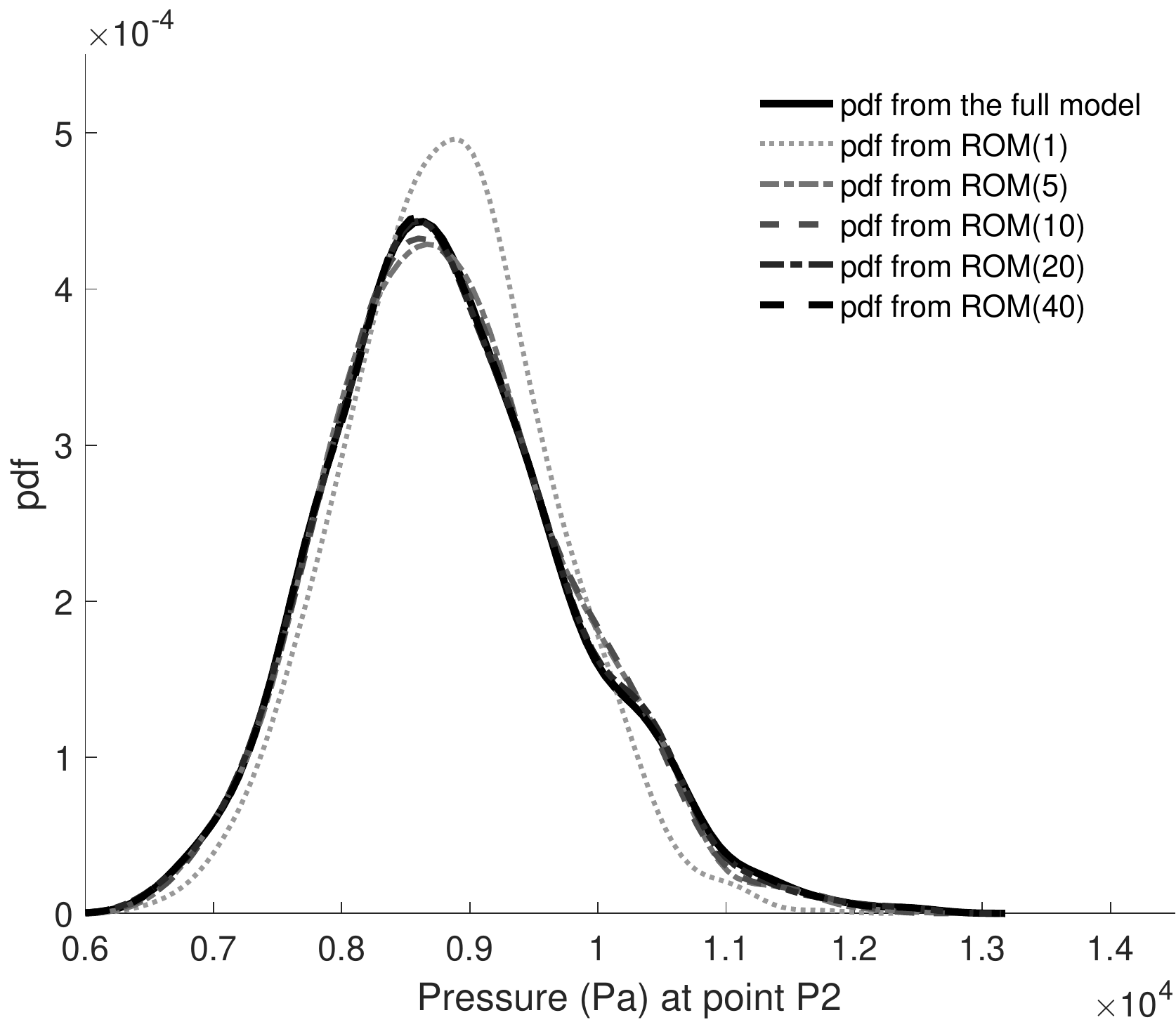}
\\
\includegraphics[width=.45\textwidth]{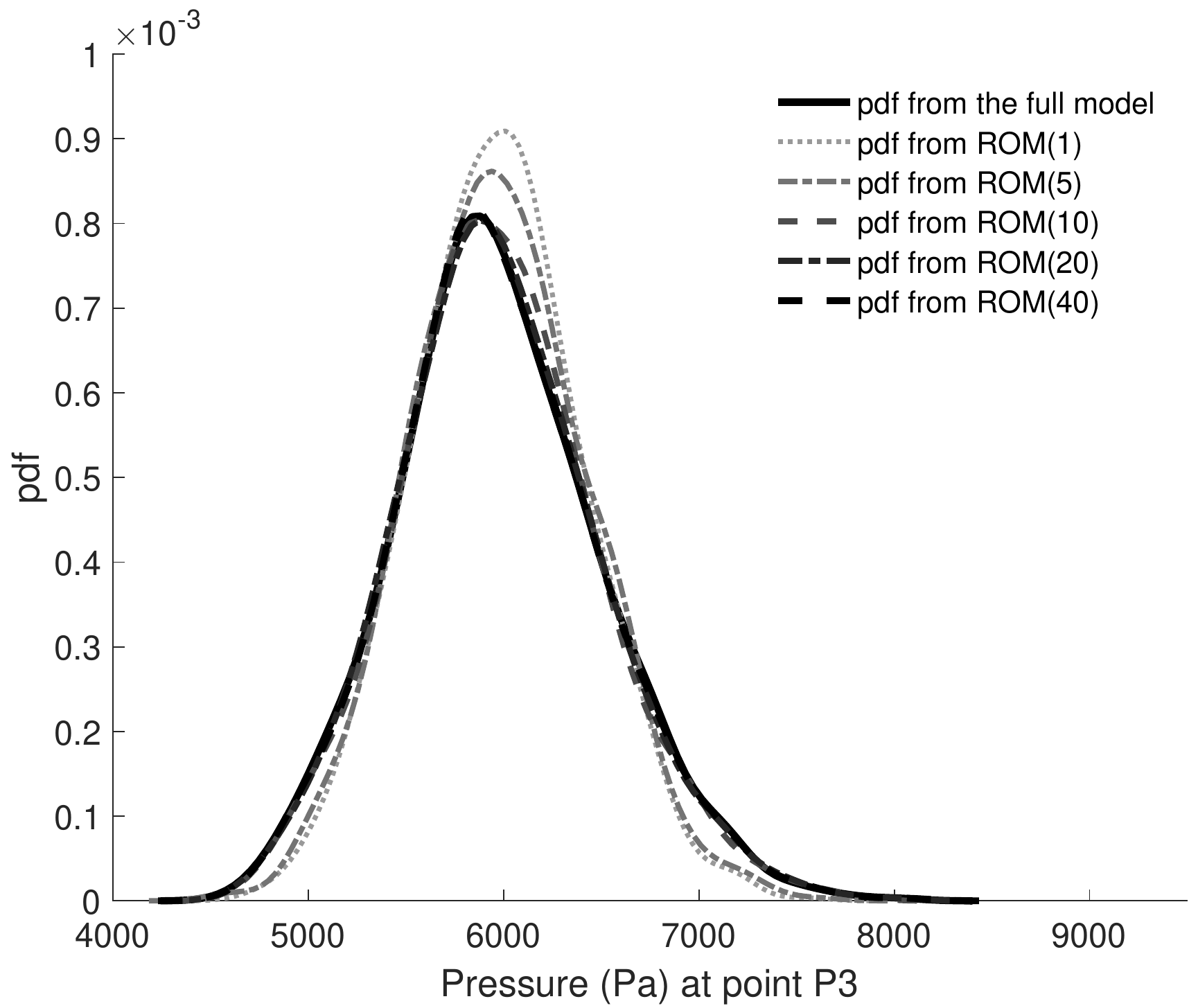}
\includegraphics[width=.45\textwidth]{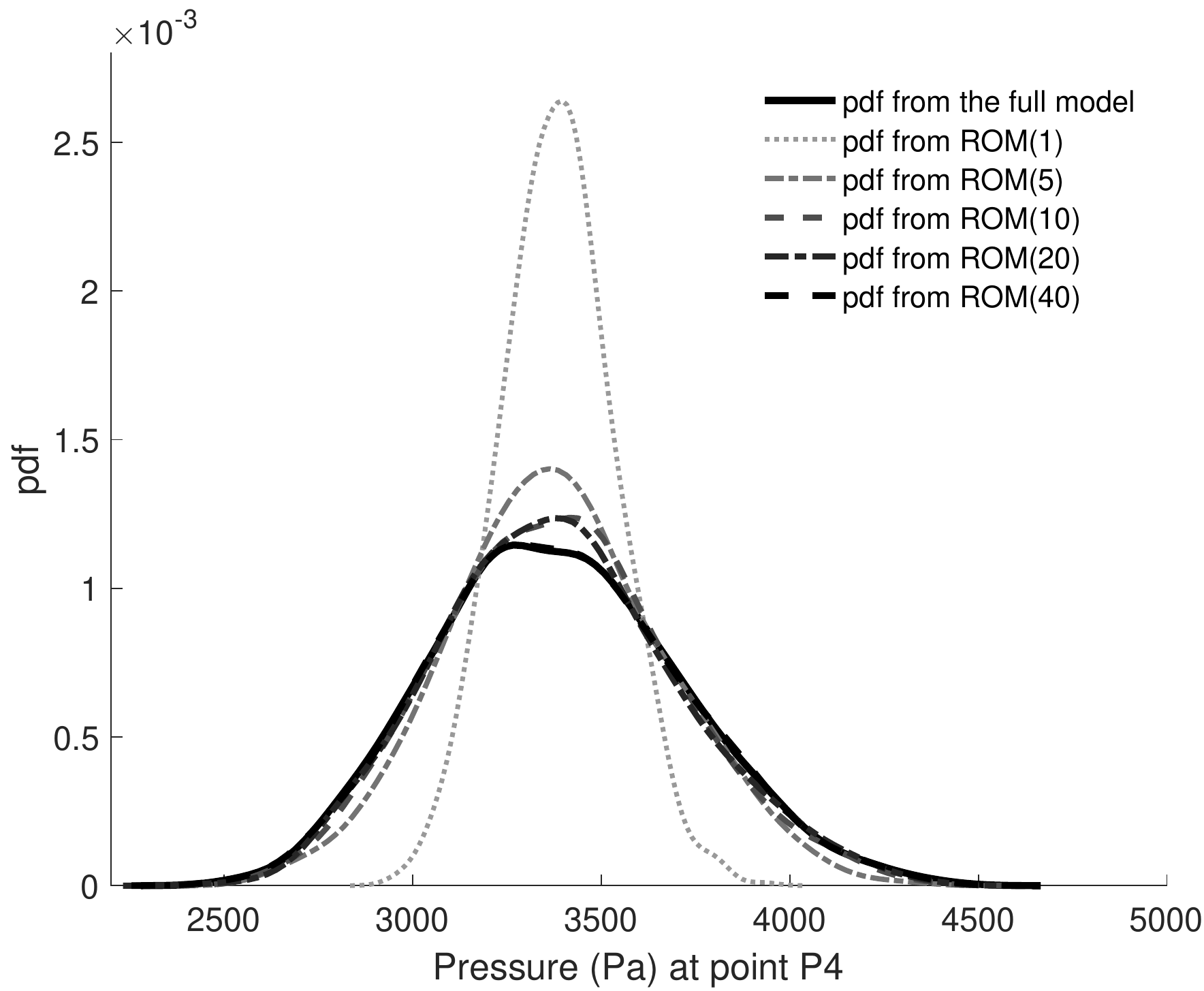}
\caption{Comparison of PDFs constructed from the output KL-based ROM with variable fidelity at points $P_1$, $P_2$, $P_3$ and $P_4$ when the correlation length $\ell$ is $0.5$ mm.}
\label{fig:PDF_0.5}
\end{figure*}

\begin{figure*}[ht]\centering
\includegraphics[width=.45\textwidth]{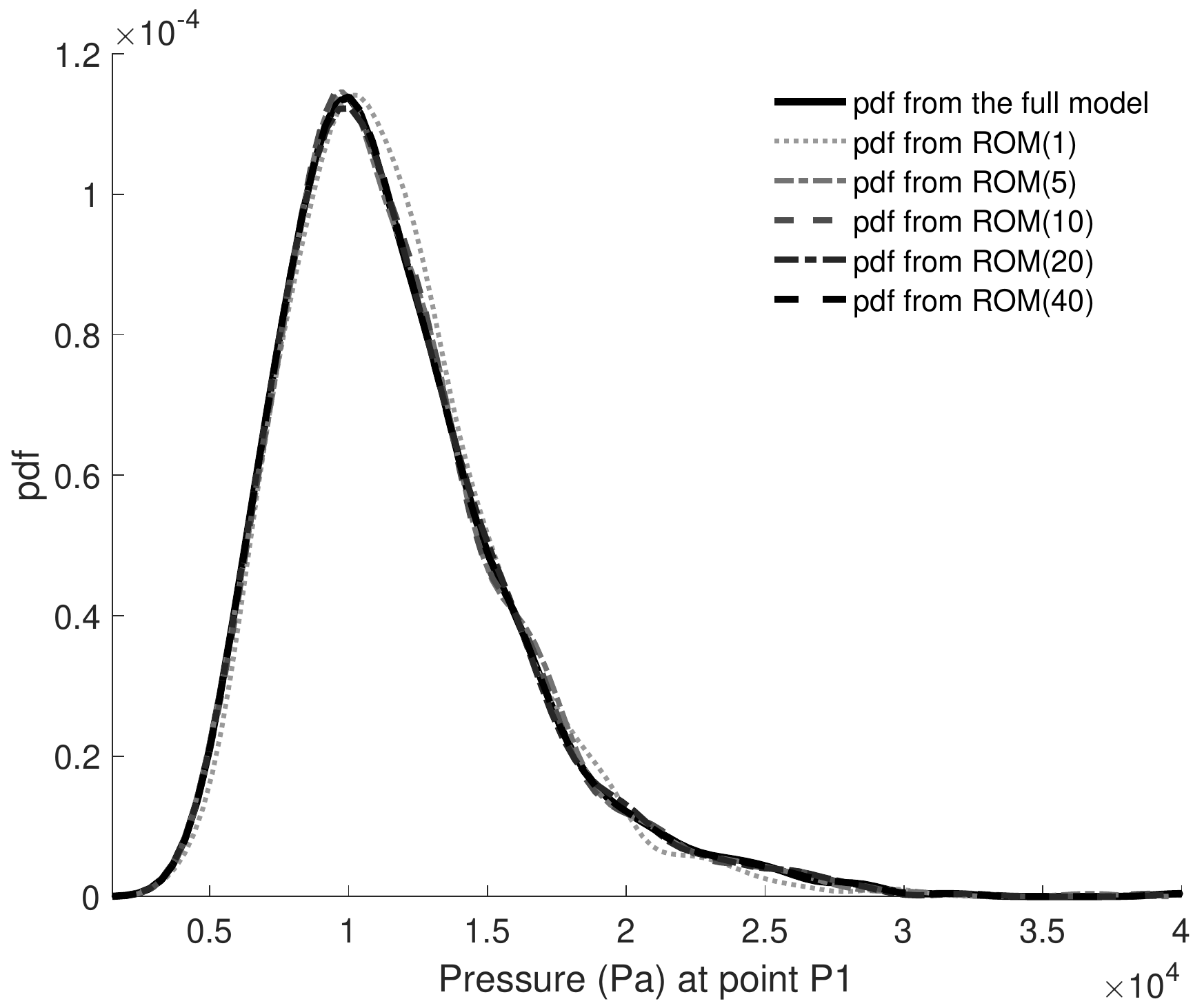}
\includegraphics[width=.45\textwidth]{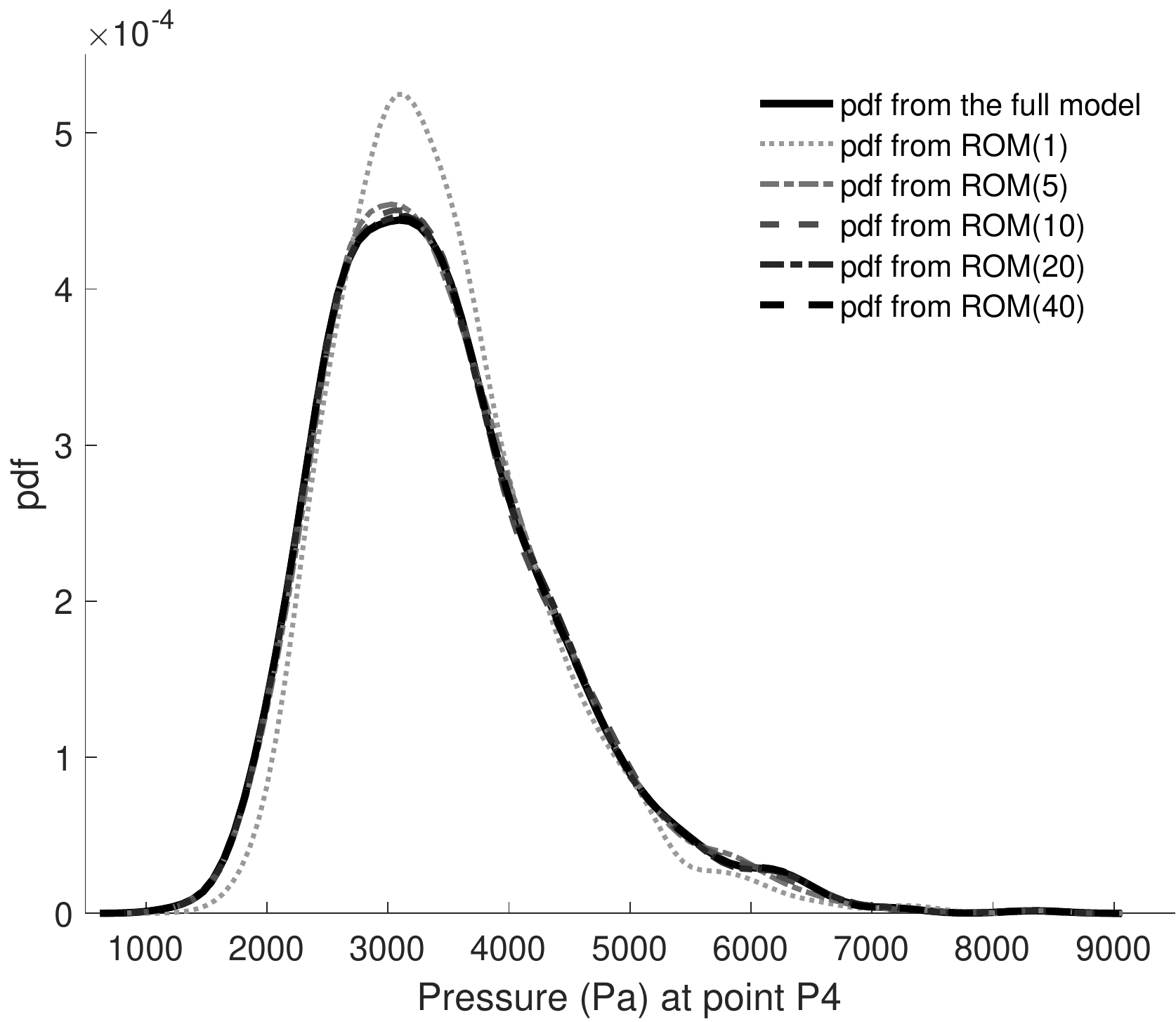}
\caption{Comparison of PDFs constructed from the output KL-based ROM with variable fidelity at points $P_1$ and $P_4$ when the correlation length $\ell$ is $3$ mm.}
\label{fig:PDF_3}
\end{figure*}

In Figure~\ref{fig:PDF_0.5}, we present the PDFs from the full model, which is
the numerical solution of the governing equation~\eqref{equ:2D_Darcy}, and
those from the ROMs with the first $n$ KL terms, where $n \in \{1, 5, 10, 20,
40\}$, when $\ell$ is $0.5$ mm. We observe that although the permeability field
in this case is very complex, almost all ROMs with the first 10 KL terms can
reasonably recover the PDFs constructed from the corresponding full models.
With the first 40 KL terms, ROMs can recover the PDFs constructed from full
models with negligible discrepancy on all the four points studied. When the
correlation length becomes larger, e.g., $\ell = 3$ mm, the effectiveness of
ROMs becomes more apparent than that with small correlation lengths. From
Figure~\ref{fig:PDF_3}, we can clearly see that when $\ell$ equals to $3$ mm,
at the point $P_1$, the ROM with only the first KL mode can almost recover the
PDF constructed from the full model; at the point $P_4$, the ROM with the first
five KL modes can capture almost all the features in the PDF.

As in the 1D model problem examined earlier, we can achieve a substantial
output dimension reduction by using the ROM given by it KL expansion.  However,
the case for the input dimension reduction is less clear in this case.  When
the correlation length $\ell$ is small, a very high input dimension is needed
to capture most of the variance in the permeability field. However, as before,
the PDE solution is still not very sensitive to high-order KL modes; for
instance, even with $\ell = 0.5$ mm, we note that the average relative error
falls below $5\%$ with an $\Nkla$ of around $50$ (see Figure~\ref{fig:approx_2d}). However, if further accuracy
is required, more input KL terms need to be retained.  A question arises: is
there a way to find a subset of the parameter KL terms that are most influential
to model output variability?  In our previous work~\cite{Cleaves19}, a
derivative-based global sensitivity approach has been established to identify
unimportant input parameters, for function-valued quantities of interest such
as the pressure field. 
The approach in~\cite{Cleaves19} guides an efficient input dimension reduction
strategy, by identifying the KL terms of the input that contribute most to
variability of the output field.

\section{CONCLUSIONS}\label{sec:conc}
We have studied the input and output dimension reduction of elliptic
PDEs, with random field input parameters, via the truncated KL expansion
technique. In this study, the covariance function of the stochastic process
defining the input parameter field is given, and that of the random output is
constructed via Monte Carlo sampling. From numerical experiments with both 1D
and 2D elliptic PDEs, we observe that when the correlation length is small,
very high-dimensional representation is needed to fully resolve the variations in
the input field. However, the elliptic operator is not sensitive to high-order
KL terms. As a result, the solution of the elliptic PDE only shows strong
dependence to the low-order KL terms of the random input field; moreover, the
eigenvalues of the solution covariance operator decay very fast.  This enables
a low-rank representation of the PDE solution in a low-dimensional input
parameter space. We then apply these dimension reduction methods in modeling
the biotransport process in tumors with uncertain material properties, and
demonstrate that the pressure field can be approximated with a low-dimensional
representation even for random permeability fields with small correlation
lengths. The efficacy of the low-rank ROMs is verified by their capability to 
recover the PDFs of the pressures at different locations in the flow field.

We demonstrate in this study that the truncated KL expansion can be an
effective approach to reduce the output dimensions of an elliptic PDE. This is
important for uncertainty quantification of large flow problems with a huge
number of spatial dimensions. Although the truncated KL expansion can also
reduce the input dimensions, its effect is not apparent when the correlation
length of the covariance function is small. Advanced dimension reduction
methods, such as global sensitivity analysis and active subspace, need to be
developed to tackle input dimension reduction. One example is our recent work
on functional derivative-base global sensitivity analysis~\cite{Cleaves19}.
More progress will be reported in our future work.   

\section*{Acknowledgments}
M.L.~Yu gratefully
acknowledge the faculty startup support from the department of mechanical
engineering at the University of Maryland, Baltimore County (UMBC).
%
%
\bibliographystyle{asmems4}
\bibliography{asme2e}

\end{document}